\documentclass[twocolumn,prd,preprintnumbers,amsmath,amssymb,nofootinbib]{revtex4}

\usepackage{graphicx}
\usepackage{dcolumn}
\usepackage{bm}
\usepackage{epsfig}
\usepackage{amssymb}
\usepackage{amsmath}

\newcommand{\aap}{Astron. Astrophys.}

\newcommand{\apjl}{Astrophys. J. Lett.}
\newcommand{\apjs}{Astrophys. J. Suppl. Ser.}

\newcommand{\jcap}{JCAP}

\newcommand{\mnras}{Mon. Not. R. Astron. Soc.}
\newcommand{\physrep}{Physics Reports}
\newcommand{\plb}{Phys. Lett. B}

\newcommand{\lsim}{\mathrel{\hbox{\rlap{\lower.55ex\hbox{$\sim$}} \kern-.3em \raise.4ex \hbox{$<$}}}}
\newcommand{\gsim}{\mathrel{\hbox{\rlap{\lower.55ex\hbox{$\sim$}} \kern-.3em \raise.4ex \hbox{$>$}}}}
\newcommand{\beq}{\begin{equation}}
\newcommand{\eeq}{\end{equation}}
\newcommand{\drm}{\mathrm{d}}

\newcommand{\gam}{\Gamma_\phi}

\newcommand{\rhos}{\rho_\phi}
\newcommand{\rhor}{\rho_r}

\newcommand{\rhom}{\rho_\chi}
\newcommand{\nchi}{n_\chi}
\newcommand{\nchieq}{n_{\chi,\mathrm{eq}}}
\newcommand{\rhoeq}{\rho_{\chi,\mathrm{eq}}}

\newcommand{\dels}{\delta_\phi}
\newcommand{\delr}{\delta_r}
\newcommand{\delm}{\delta_\chi}
\newcommand{\deleq}{\delta_{\chi,\mathrm{eq}}}
\newcommand{\thes}{\theta_\phi}
\newcommand{\ther}{\theta_r}
\newcommand{\them}{\theta_\chi}

\newcommand{\tilk}{\tilde{k}}
\newcommand{\kdec}{k_\mathrm{RH,h}}
\newcommand{\krh}{k_\mathrm{RH}}

\newcommand{\adec}{a_\mathrm{RH}}

\newcommand{\mdec}{M_\mathrm{RH}}

\newcommand{\delcoll}{\delta_c}
\newcommand{\avgE}{\langle E_\chi \rangle}
\newcommand{\sigv}{\langle \sigma v \rangle}
\newcommand{\gstarRH}{g_{*\mathrm{RH}}}
\newcommand{\Trh}{T_\mathrm{RH}}
\newcommand{\mpl}{m_\mathrm{Pl}}
\newcommand{\mdm}{m_\chi}
\newcommand{\sigunits}{\,\mathrm{cm^3\,s^{-1}}}
\newcommand{\kcut}{k_\mathrm{cut}}
\newcommand{\Tkds}{T_\mathrm{kdS}}
\newcommand{\Tkd}{T_\mathrm{kd}}
\newcommand{\kkd}{k_\mathrm{kd}}
\newcommand{\kfs}{k_\mathrm{fs}}

\begin{document}

\title{The Dark Matter Annihilation Boost from Low-temperature Reheating}

\author{Adrienne L. Erickcek}\email{erickcek@physics.unc.edu}
\affiliation{Department of Physics and Astronomy, University of North Carolina at Chapel Hill, Phillips Hall CB 3255, Chapel Hill, NC 27599 USA}


\begin{abstract}
The evolution of the Universe between inflation and the onset of Big Bang Nucleosynthesis is difficult to probe and largely unconstrained.  This ignorance profoundly limits our understanding of dark matter: we cannot calculate its thermal relic abundance without knowing when the Universe became radiation dominated.  Fortunately, small-scale density perturbations provide a probe of the early Universe that could break this degeneracy.  If dark matter is a thermal relic, density perturbations that enter the horizon during an early matter-dominated era grow linearly with the scale factor prior to reheating.  The resulting abundance of substructure boosts the annihilation rate by several orders of magnitude, which can compensate for the smaller annihilation cross sections that are required to generate the observed dark matter density in these scenarios.  In particular, thermal relics with masses less than a TeV that thermally and kinetically decouple prior to reheating may already be ruled out by Fermi-LAT observations of dwarf spheroidal galaxies.  Although these constraints are subject to uncertainties regarding the internal structure of the microhalos that form from the enhanced perturbations, they open up the possibility of using gamma-ray observations to learn about the reheating of the Universe.    
\end{abstract}
 
\maketitle

\section{Introduction}

The expansion history of the Universe prior to the onset of Big Bang Nucleosynthesis (BBN) is unknown, but the nearly scale-invariant spectrum of the primordial curvature perturbations provides strong evidence that the Universe experienced a period of inflation \cite{Guth80, AS82, Linde82} shortly after the Big Bang.  Inflation leaves the Universe devoid of radiation, and yet the primordial abundance of light elements indicates that the Universe was radiation dominated during BBN.   Therefore, the Universe must have become radiation dominated at some point after inflation, but we do not know when that transition occurred, nor do we know how the Universe evolved between inflation and the onset of radiation domination.  The existence of the cosmic neutrino background provides the most robust constraint on the temperature of the Universe when it became radiation dominated: this reheat temperature must exceed 3 MeV to generate the neutrinos required to produce the observed abundances of light elements \cite{KKS99, KKS00, Han04, IKT05} and the observed power spectra of anisotropies in the cosmic microwave background and large-scale density perturbations \cite{IKT07,dBPM08}.  The vast difference between 3 MeV and the theorized energy scale of inflation leaves a tremendous gap in our understanding of the thermal history of the Universe.  

In the simplest models, inflation is powered by a single scalar field, the inflaton.  Inflation ends when the inflaton begins to oscillate around the minimum of its potential, and the Universe becomes radiation dominated when the inflaton eventually decays into relativistic particles  \citep{ASTW82, Turner83, TB90, KLS94, KLS97, DFKPP06, ABCM10, MR10}.  If the potential is quadratic, a coherently oscillating scalar field has the same dynamics as a pressureless fluid \cite{Turner83}.  Therefore, most inflationary scenarios include a transient period of effective matter domination between the end of inflation and radiation domination.  Furthermore, after the inflaton decays into radiation, other oscillating scalar fields may come to dominate the energy density of the Universe; such scalars commonly result from stabilized moduli in string theories \cite{1983PhLB..131...59C, CCQR93, BKN94,1995PhRvD..52..705B, 1995PhRvD..52.3548B} and are responsible for generating the primordial curvature perturbation in the curvaton model \cite{Mollerach90, LM97, LW02, MT01}.  It is also possible that quasi-stable massive particles could dominate the Universe prior to BBN \cite[e.g.][]{Zhang15}.  Thus, a pre-BBN early-matter-dominated era (EMDE), driven by either an oscillating scalar field or a quasi-stable massive particle, is a generic prediction of several early-Universe theories.  

There may have been several EMDEs, as the energy density of longer-lived oscillating scalar fields could surpass the energy density of the relativistic decay products generated at the end of earlier EMDEs.  In this case, the last EMDE is most cosmologically significant because it generates the current content of the Universe and its end determines the timing of reheating, which is defined to be the onset of the final radiation-dominated era.  Reheating is usually assumed to occur too early to have any cosmological impact, but there are strong motivations to consider lower reheat temperatures.  The moduli fields predicted by string theory have gravitational couplings, so their masses must be $\gsim$10 TeV to avoid violating the BBN bound on the reheat temperature ($\Trh \gsim 3$ MeV).  String theories also generally predict that the lightest modulus field should not be significantly more massive than the gravitino, which implies that the modulus mass should be less than $\sim$1000 TeV if supersymmetry is to mitigate the electroweak hierarchy problem.  Consequently, the reheat temperature in these theories is generally less than a few hundred GeV (see Ref. \cite{KSW15} for a recent review).  

Our ignorance of the pre-BBN thermal history profoundly limits our understanding of the origins of dark matter \cite[e.g.][]{KT90, GKR01, GG06, DIK06, GSK08, Watson10}, the origins of the baryon asymmetry \citep[e.g.][]{GKR01, ADS10, DMST2010}, and the connection between the primordial power spectrum and the inflaton potential \cite[e.g.][]{DKW14, 2015PhRvL.114h1303M,2015PhRvD..91d3521M}.  In particular,  the relic abundance of dark matter is profoundly altered if dark matter thermally decouples during an EMDE \cite{KT90, CKR99, GKR01, FRS03, Pallis04, GG06, GGSY06, Drewes14, RTT14, KKN15}; the decaying field dilutes the relic abundance of thermal dark matter, so a smaller value of $\sigv$ is required to generate the observed dark matter density.  The dark matter density may be enhanced, however, if the dark matter is also produced nonthermally (as a decay product, for instance),
in which case $\sigv$ may need to be increased so that the excess dark matter is eliminated.  Ref. \cite{GGSY06} demonstrated that nearly any supersymmetric dark matter particle, with a wide range of $\sigv$ values, can give the correct relic abundance for some combination of a low reheat temperature and nonthermal production.  Even in the absence of nonthermal production, it is possible to obtain the observed density of dark matter if $\sigv \ll 3\times10^{-26}$ cm$^{3}$ s$^{-1}$.  Therefore, constraints on the dark matter annihilation rate cannot rule out a thermal origin for dark matter as long as the reheat temperature remains unconstrained.

In this work, I show that an EMDE's effect on density perturbations on scales that enter the cosmological horizon prior to reheating can significantly enhance the dark matter annihilation rate by increasing the abundance of microhalos.  During an EMDE, subhorizon perturbations in the dominant energy density grow linearly with the scale factor \citep{JLM10a, EFG10}.   If the duration of the EMDE is sufficiently long, these perturbations become nonlinear and could produce gravitational waves \cite{AW09, JLM10b}, but probably not black holes \cite{KSW15}.   Ref. \cite{ES11} demonstrated that nonrelativistic decay products produced during the EMDE inherit these enhanced perturbations.  Consequently, if dark matter is generated nonthermally and is nonrelativistic at reheating, perturbations in the dark matter density that enter the cosmological horizon during the EMDE are significantly enhanced compared to larger-scale perturbations.  Ref. \cite{ES11} showed that this small-scale inhomogeneity would radically increase the number density of sub-Earth-mass microhalos.  Unfortunately, these microhalos are too small to detect with gravitational lensing in the near future \cite{LEM12}.  This analysis was later extended to scenarios in which the dark matter is generated nonthermally and then annihilates down to the observed abundance \cite{BR14, FOW14}.  In these scenarios, subhorizon perturbations in the dark matter density grow linearly during the EMDE and then decrease slightly at reheating.  Although this brief decrease does not erase the growth of the perturbations during the EMDE, it is likely that density perturbations on these small scales are further suppressed by free-streaming of the dark matter particle after reheating \cite{FOW14}.  

However, if the dark matter is produced thermally during an EMDE, it may kinetically decouple prior to reheating, in which case the free-streaming scale is far smaller than the horizon scale at reheating \cite{GG08}.  Furthermore, the dark matter's thermal origins demand that it must annihilate in high-density environments, which provides a means of detecting the microhalos generated by the EDME.  This scenario is the focus of this work; I will assume that the dominant energy component during the EMDE either does not decay into dark matter or that the branching ratio for decay into dark matter is small enough that the generated dark matter density is far smaller than the thermal relic density.  I will show that subhorizon perturbations in the thermally generated dark matter density grow linearly during EMDE and that the resulting matter power spectrum is identical to the power spectrum derived in Ref. \cite{ES11} for nonthermal dark matter in the absence of annihilations.  

If the dark matter kinetically decouples prior to reheating, the largest scales enhanced by the EMDE are not suppressed by the free-streaming of the dark matter particle nor by its elastic scatterings with relativistic particles.  Therefore, these enhanced small-scale perturbations will significantly increase the abundance of microhalos: instead of being rare, isolated objects, microhalos contain most of the dark matter in these scenarios.  Even though these particles must have annihilation cross sections that are far smaller than the canonical thermal cross section, the boost to the annihilation rate from the microhalos is sufficient in some cases to increase the annihilation rate beyond current observational limits from the \emph{Fermi Gamma-Ray Telescope} \cite{FermidSphs15, FermiIGRB15}.  This result demonstrates that gamma-ray observations are capable of probing the evolution of the Universe prior to BBN and constraining the origins of dark matter.

I begin in Section \ref{sec:abundance} by reviewing the thermal production of dark matter during an EMDE.  In Section \ref{sec:pertevol}, I analyze the evolution of density perturbations during the EMDE and after reheating, recovering the matter power spectrum found in Ref. \cite{ES11}.  The suppression of small-scale perturbations due to elastic scattering and free-streaming of the dark matter particles is considered in Section \ref{sec:cut}.  This section also explores the impact the decoupling scale has on the microhalo population.  In Section \ref{sec:boost}, I estimate how these microhalos boost the dark matter annihilation rate, and I compare the resulting annihilation rate to constraints from gamma-ray observations in Section \ref{sec:detect}.  Finally, I summarize my results and discuss avenues for future investigation in Section \ref{sec:con}.  The appendices present the derivation of the perturbation evolution equations and their initial conditions.  Natural units ($\hbar=c=k_B=1$) are used throughout this work. 

\section{Thermal Dark Matter during an Early Matter-Dominated Era}
\subsection{Background Evolution}
\label{sec:abundance}
The dominant component of the Universe prior to reheating is modeled as a pressureless fluid that decays perturbatively into relativistic particles. Although I will refer to this pressureless fluid as a scalar field, it could equally well be a collection of massive particles.  I assume that dark matter is created thermally and that no dark matter is produced during the decay of the scalar field.  In this scenario, the background equations for the energy density $\rhos$ of the scalar field, the radiation density $\rhor$ (not including dark matter particles, even if they are relativistic), and the number density of dark matter particles $\nchi$ are
\begin{subequations}
\begin{align}
\frac{\drm}{\drm t} \rhos+3H\rhos &= -\gam \rhos, \label{bkgdS}\\
\frac{\drm}{\drm t} \rhor+4H\rhor &= \gam \rhos + {\sigv} \avgE \left( \nchi^2 - \nchieq^2 \right),\label{bkgdR}\\
\frac{\drm}{\drm t} \nchi+3H\nchi&= - {\sigv}\left( \nchi^2 - \nchieq^2 \right),  \label{bkgdM}
\end{align}
\label{bkgd}%
\end{subequations}
where $\gam$ is the scalar decay rate; ${\sigv}$ is the velocity-averaged cross section for dark matter annihilations; $\avgE = \rhom/\nchi$ is the average energy of a dark matter particle; and $\nchieq$ is the number density of dark matter particles in thermal equilibrium.\footnote{These equations differ slightly from those in Ref. \cite{GKR01}, which includes an extra factor of two in the radiation equation to account for the fact that two photons are produced during each annihilation event.  This factor of two is not necessary if the dark matter is composed of Majorana particles, because then the rate of annihilation events per volume is ${\sigv}\nchi^2/2$.  However, its inclusion does not affect the relic abundance of dark matter.} 
For a dark matter particle with mass $m_\chi$ and $g_\chi$ internal degrees of freedom,
\beq
\nchieq =  \frac{g_\chi}{2\pi^2} \int_{m_\chi}^\infty \frac{\sqrt{E^2-m_\chi^2}}{e^{E/T}+1} E \,\drm E,
\eeq
where $T$ is the temperature of the radiation bath.  When the dark matter particle is nonrelativistic, $\avgE \simeq m_\chi$, while $\avgE \simeq 3.15 T$ if $m_\chi \ll T$.  To facilitate a numerical solution to the background equations, I make the approximation that $\avgE \simeq \sqrt{m_\chi^2 + 9.93 T^2}$, which matches  $\rhom/\nchi$ to within 10\%.  This system of equations was studied extensively in Ref. \cite{GKR01}; in this section I will only summarize the key results that are required to compute the relic abundance of dark matter in these scenarios.

\begin{figure}
 \centering
 \resizebox{3.4in}{!}
 {
      \includegraphics{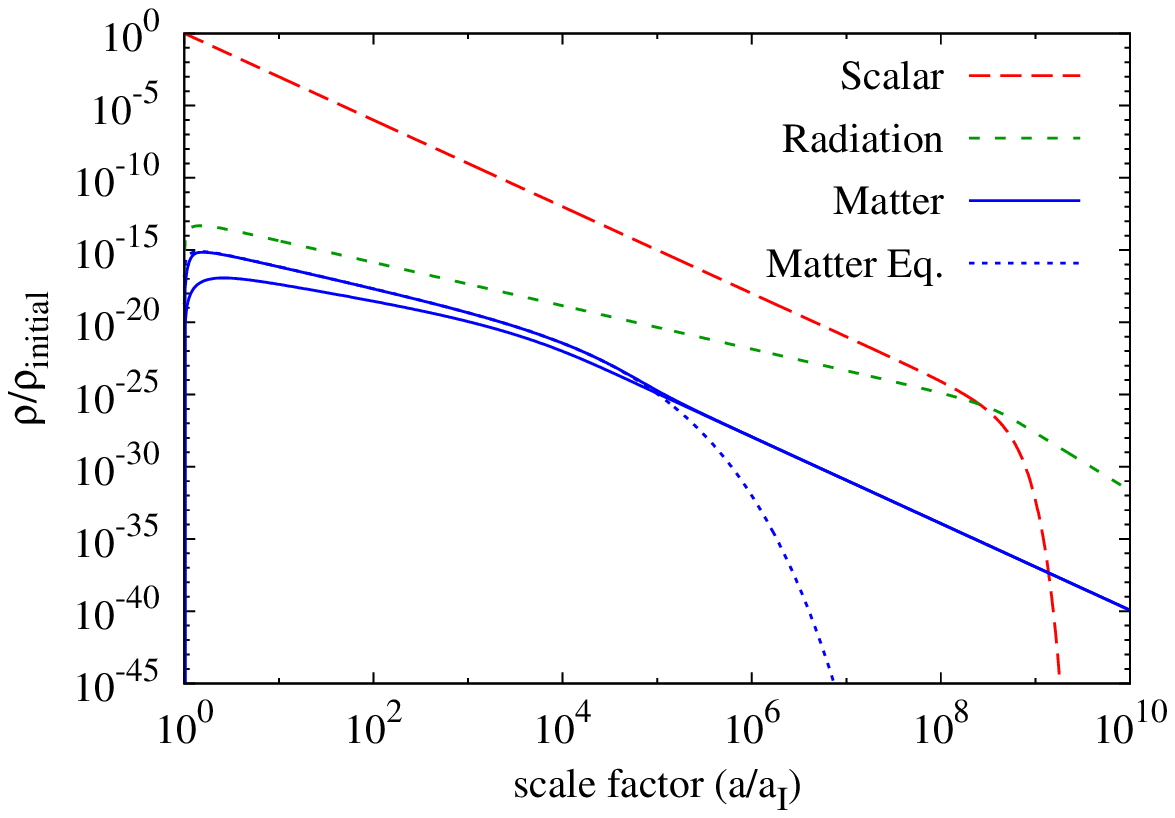}
 }
\caption{The densities of the decaying scalar field, radiation, and matter during reheating.  In this figure, $m_\chi = 750$ GeV, and $H=\gam$ when $a/a_I=2\times10^8$, which corresponds to a reheat temperature of 5 GeV.  The two solid curves show $\rhom$ for the two values of $\sigv$ that generate the dark matter density observed today ($\Omega_\chi h^2 = 0.12$) for these values of $\mdm$ and $\Trh$.  The top solid curve shows a freeze-out scenario with $\sigv = 9.0\times10^{-31} \mathrm{cm^3\,s^{-1}}$, whereas the bottom solid curve shows a freeze-in scenario with $\sigv = 2.9 \times 10^{-33} \mathrm{cm^3\,s^{-1}}$.  The dotted curve shows the equilibrium density of dark matter, $\rho_\mathrm{eq} = \avgE \nchieq.$}
\label{Fig:bkgd}
\end{figure}

Figure \ref{Fig:bkgd} shows the evolution of the densities determined by solving Eq.~(\ref{bkgd}).  The scalar field dominates the energy content of the Universe until $\gam \simeq H$.  At that time, $\rhos$ begins to decay exponentially, and the Universe becomes radiation dominated.  Although radiation does not instantaneously dominate the energy density when $\gam = H$, it is still useful to define the reheat temperature $\Trh$ by the relation 
\beq
\gam =  \sqrt{\frac{8\pi^3 g_*(\Trh)}{90}}\frac{\Trh^2}{\mpl},
\label{Trhdef}
\eeq
where $\mpl = \sqrt{1/G}$ is the Planck mass, and \mbox{$g_*(T) \equiv \rhor(T)/[(\pi^2/30)T^4]$} is the number of relativistic degrees of freedom.  I evaluate $g_*(T)$ by summing the contributions of all Standard Model particles to $\rho_r$ (see Appendix A of Ref.~\cite{EBB14}).  This definition of the reheat temperature can be used to eliminate $\gam$ from Eq.~(\ref{bkgd}); $\Trh$ then parametrizes the onset of radiation domination.  It is also useful to define $\adec$ to be the value of the scale factor when $\gam = H$.  Since $H \propto a^{-3/2}$ during the EMDE, 
\beq
\frac{\adec}{a_I} = \left[ \frac{30\rho_{\phi,I}}{\pi^2 \gstarRH \Trh^4}\right]^{1/3},
\eeq
where $a_I$ is the value of the scale factor at some point during scalar domination, $\rho_{\phi,I}$ is $\rhos$ when $a=a_I$, and I have introduced the shorthand \mbox{$\gstarRH \equiv g_*(\Trh)$}.  In Figure \ref{Fig:bkgd}, $\Trh= 5$ GeV and $\adec/a_I = 2\times 10^{8}$. 

\begin{figure*}[t]
 \centering
\begin{minipage}{0.5\textwidth}
\centering
 \resizebox{3.4in}{!}
 {
      \includegraphics{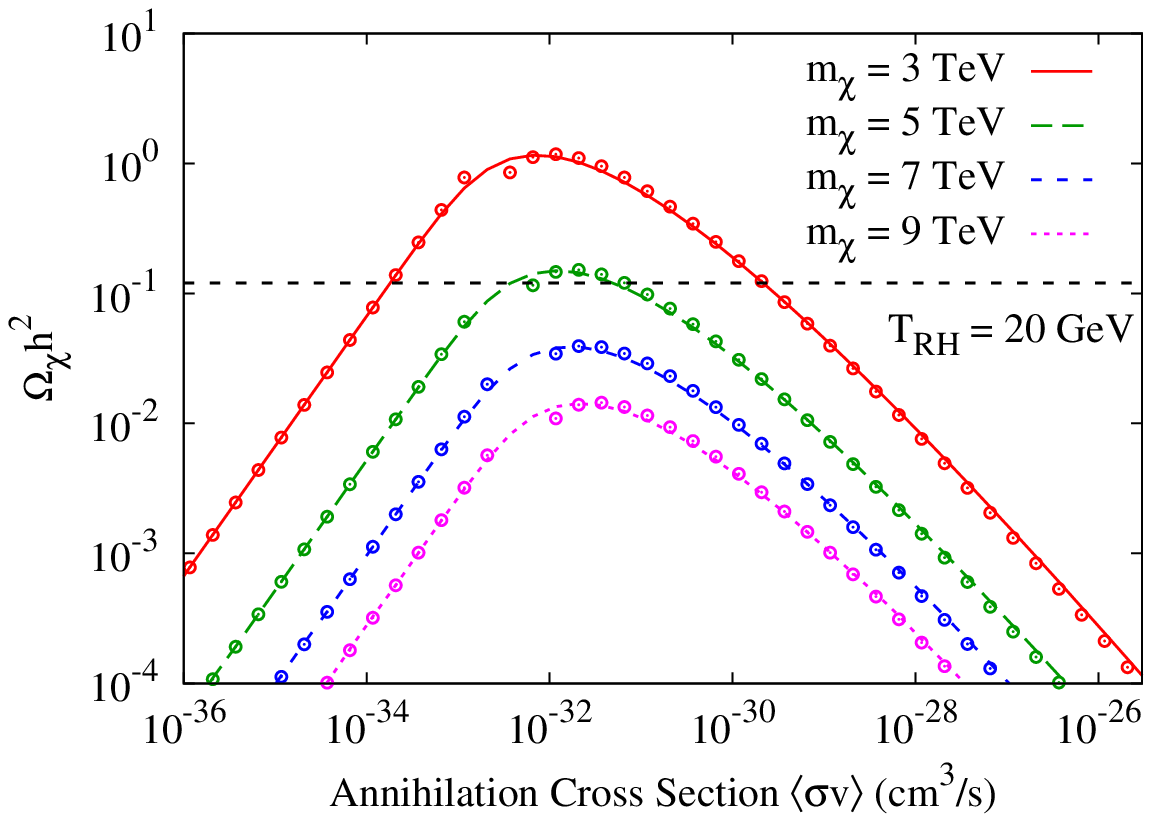}
 }
\end{minipage}%
\begin{minipage}{0.5\textwidth}
\centering
 \resizebox{3.4in}{!}
{
      \includegraphics{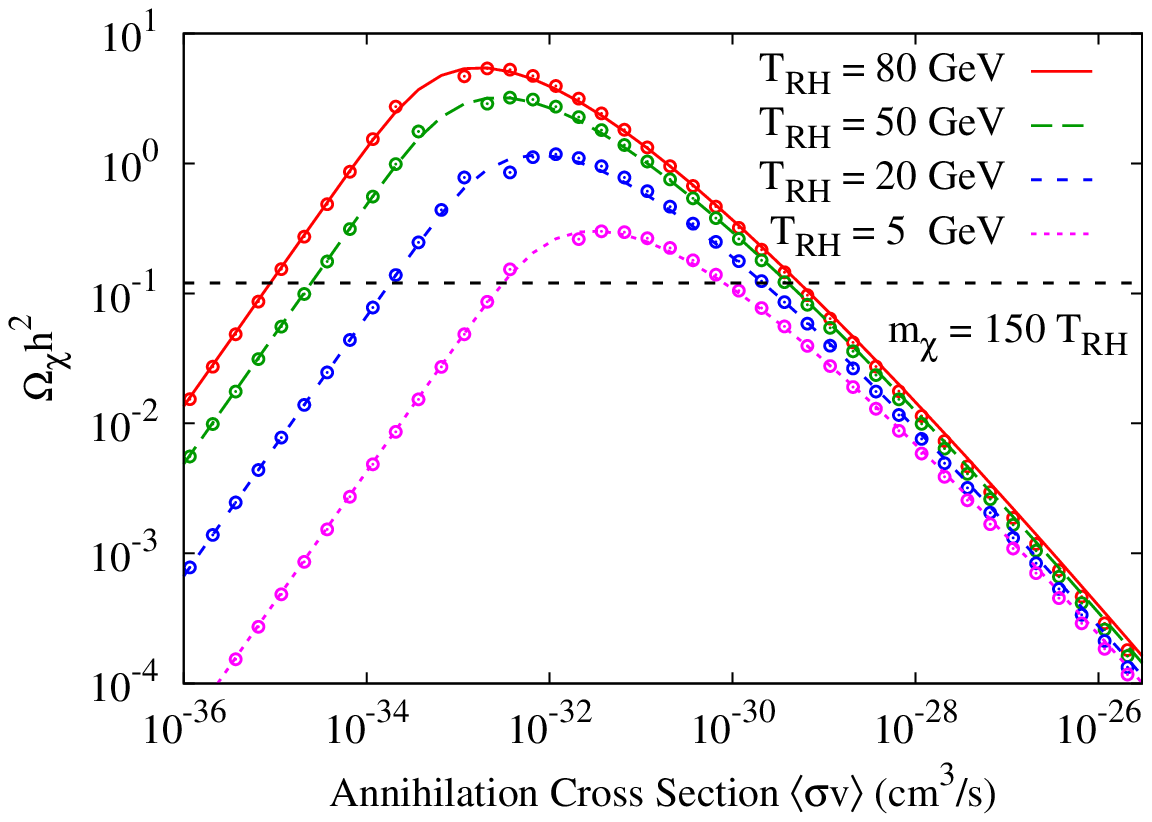}
 }
\end{minipage}%
\caption{The present-day abundance of dark matter $\Omega_\chi h^2$ as a function of the velocity-averaged annihilation cross section $\sigv$.  For small $\sigv$ values, the dark matter ``freezes in" and $\Omega_\chi h^2 \propto \sigv$, whereas for larger $\sigv$ values, the dark matter ``freezes out" and $\Omega_\chi h^2 \propto \sigv^{-1}$.  The left panel shows that increasing $\mdm/\Trh$ decreases $\Omega_\chi h^2$ in both cases.  The right panel shows that decreasing $\Trh$ while keeping $\mdm/\Trh$ fixed also decreases $\Omega_\chi h^2$, albeit mildly if $\sigv$ is in the freeze-out regime.  In both panels, the curves are based on numerical solutions to Eq.~(\ref{bkgd}), while the open circles show the predictions of Eqs.~(\ref{OmegaFO}) and (\ref{OmegaFI}).  The dashed line marks the observed value established by the \emph{Planck} mission: $\Omega_\chi h^2 = 0.12$ \cite{2015planck}.}
\label{Fig:OmegaDM}
\end{figure*}

Prior to the onset of radiation domination, the transfer of energy from the scalar field to radiation forces \mbox{$\rhor \propto a^{-3/2}$}.  This scaling can be derived from Eq.~(\ref{bkgdR}) in the limit that $ \gam \rhos$ dominates over all other contributions to $\drm \rhor/\drm t$; since $\rhos \propto a^{-3}$ and $H \propto a^{-3/2}$ during the EMDE, Eq.~(\ref{bkgdR}) then implies that $\rhor'(a) \propto a^{-5/2}$.  To obtain an explicit expression for $\rhor$ during the EMDE, however, the $4H\rhor$ term in Eq.~(\ref{bkgdR}) must be included, and an initial value of $\rhor$ must be provided.  If $\rhor = 0$ when $a=a_I$, then
\beq
\rhor = \frac{2}{5}\Trh^2  \sqrt{\frac{\pi^2 \gstarRH \rho_{\phi,I}}{30}} \left[\left(\frac{a}{a_I}\right)^{-3/2}-\left(\frac{a}{a_I}\right)^{-4}\right].
\label{rhorad}
\eeq
As shown in Figure \ref{Fig:bkgd}, $\rhor$ increases rapidly after $a=a_I$, reaches a maximum when $a/a_I \simeq 1.48$, and then decreases as $\rhor \propto a^{-3/2}$ until the onset of radiation domination. Since the radiation cools during the EMDE ($T\propto a^{-3/8}$), $\Trh$ is \emph{not} the maximum temperature of the radiation bath, and it is possible to thermally produce particles with $m_\chi \gg \Trh$.  Dark matter is efficiently produced provided that $m_\chi < T_\mathrm{MAX}$, where
\beq
\frac{T_\mathrm{MAX}}{\Trh} \simeq 0.5 \frac{\gstarRH^{1/8}}{g_*(T_\mathrm{MAX})^{1/4}} \frac{[\mpl H(a_I)]^{1/4}}{\Trh^{1/2}}.
\eeq
As long as $T_\mathrm{MAX} \gsim 2 m_\chi$, the relic abundance of dark matter will be independent of $T_\mathrm{MAX}$, which implies that it is also independent of $H(a_I)$.  Throughout this analysis, I will chose $\rho_{\phi,I}$ such that $T_\mathrm{MAX} = 5 m_\chi$, which ensures that the results are not sensitive to the details of the onset of the scalar decay.  

There are two thermal dark matter production scenarios illustrated in Figure \ref{Fig:bkgd} that both produce the same present-day dark matter density.  If $\sigv$ is sufficiently large that pair production brings the dark matter into thermal equilibrium, then $\nchi = \nchieq$ while $H < \sigv \nchieq$.  When $H \simeq \sigv \nchieq$,  the dark matter ``freezes out," and $\nchi \propto a^{-3}$ thereafter.   If dark matter freezes out prior to reheating, 
\begin{align}
\Omega_\chi h^2 \simeq \,\,& 1.6 \times 10^{-4}\, \frac{\sqrt{\gstarRH}}{g_*(T_f)} \left( \frac{\mdm/T_f}{15}\right)^4\left( \frac{150}{\mdm/\Trh}\right)^3 \nonumber \\
&\times\left(\frac{3 \times 10^{-26} \sigunits}{\sigv}\right), \label{OmegaFO}
\end{align}
where $T_f$ is the freeze-out temperature defined by \mbox{$H(T_f) = \sigv \nchieq(T_f)$}.  The ratio $\mdm/T_f$ depends only logarithmically on $\sigv$, $\Trh$, $g_\chi$, and $\mdm$; in most cases $5 \lsim \mdm/T_f \lsim 25$.  Equation~(\ref{OmegaFO}) indicates that increasing $\mdm/\Trh$ dramatically reduces the relic abundance of dark matter, as illustrated in Figure \ref{Fig:OmegaDM}.  Since $\mdm/T_f$ is nearly constant, increasing $\mdm/\Trh$ is roughly equivalent to decreasing $\Trh/T_f$.  The resulting decrease in $\Omega_\chi h^2$ has a simple explanation: if the dark matter freezes out during the EMDE, then the subsequent transfer of energy from the scalar to radiation will increase the number of photons per dark matter particle, which dilutes the relic abundance of dark matter.  As shown in Figure \ref{Fig:OmegaDM}, the observed abundance of dark matter ($\Omega_\chi h^2 = 0.1197 \pm 0.0022$ \cite{2015planck}) can only be generated if $\sigv \ll 3 \times 10^{-26} \sigunits$.

As $\sigv$ decreases, the relic abundance of dark matter increases, until $\sigv$ becomes small enough that pair production cannot bring $\nchi$ up to its equilibrium value before pair production ceases when $T \lsim \mdm/4$.  At that point, the dark matter ``freezes in," and the comoving number density of dark matter is subsequently preserved, as demonstrated in Figure \ref{Fig:bkgd}.   If the dark matter becomes nonrelativistic prior to reheating, the freeze-in abundance of dark matter is 
\begin{align}
\Omega_\chi h^2 \simeq \,\,& 0.062\, \frac{\gstarRH^{3/2}}{[g_*(\mdm/4)]^3}\left(\frac{g_\chi}{2}\right)^2 \left( \frac{150}{\mdm/\Trh}\right)^5 \nonumber \\
& \times\left(\frac{\Trh}{5\,\mathrm{GeV}}\right)^2 \left( \frac{\sigv}{10^{-36} \sigunits}\right).
\label{OmegaFI}
\end{align}
In contrast to the freeze-out scenario, where increasing $\sigv$ lowers the dark matter abundance by making dark matter annihilation more efficient, annihilations are not common in the freeze-in scenario, and increasing $\sigv$ increases the relic abundance by making pair production more efficient.  However, increasing $\mdm/\Trh$ still decreases the relic abundance in the freeze-in regime for the same reason as before: as $\mdm/\Trh$ increases, more photons are created after the production of dark matter halts.   

The final abundance of dark matter in both the freeze-in and freeze-out scenarios was derived in Ref. \cite{GKR01} using the approximation that the transition from matter domination to radiation domination occurs instantly when $T = \Trh$.  These analytic estimates accurately recover the functional dependence of the dark matter density on $\Trh$, $\mdm$, and $\sigv$, but their normalization must be adjusted slightly to correct for the fact that reheating is not instantaneous.  The expressions for $\Omega_\chi h^2$ given by Eqs.~(\ref{OmegaFO}) and (\ref{OmegaFI}) have been normalized to match the dark-matter density obtained by numerically solving Eq.~(\ref{bkgd}).  Figure \ref{Fig:OmegaDM} demonstrates how well these expressions agree with the numerical results for a wide range of parameters.
 
\subsection{Perturbation Evolution}
\label{sec:pertevol}
An EMDE also affects the evolution of the density perturbations in both the radiation and the dark matter \cite{ES11, BR14, FOW14}.   
The evolution of perturbations in the matter density is primarily determined by the expansion rate of the Universe.  While the Universe is radiation dominated, the fractional perturbation in the matter density grows logarithmically with the scale factor after the perturbation mode enters the horizon ($k > aH$, where $k$ is the wave number of the plane-wave perturbation).  When the Universe becomes matter dominated, subhorizon perturbations begin to grow linearly with the scale factor.  An EMDE has the same effect on the dark matter density even though a different pressureless fluid dominates the Universe.  If dark matter is generated during scalar decays and does not annihilate, density perturbations in the dark matter on subhorizon scales grow linearly with the scale factor during an EMDE, and then they smoothly transition to logarithmic growth when the Universe becomes radiation dominated \cite{ES11}.  

Subhorizon density perturbations in the scalar field also grow linearly with the scale factor during the EMDE.  The relativistic decay products of the scalar field inherit this inhomogeneity; subhorizon perturbations in the radiation density grow during an EMDE, but that growth is lost when the Universe becomes radiation dominated \cite{ES11}.  When dark matter annihilations are added to scenarios in which dark matter is primarily produced during scalar decays, the dark matter perturbations still grow linearly during the EMDE, but they decrease slightly when much of the dark matter annihilates during the transition to radiation domination \cite{BR14, FOW14}.   

In this section I explore how an EMDE affects perturbations if the dark matter is not produced in scalar decays and thermally decouples prior to reheating.  Evolution equations for the fractional density perturbations in $\rhos$, $\rhom$, and $\rhor$ [\mbox{$\delta_i \equiv (\rho_i  - \rho_i^0)/\rho_i^0$}, where $\rho_i^0(t)$ is the spatially uniform background density] are obtained by perturbing the covariant form of the energy-transfer equations given in Eq.~(\ref{bkgd}).   This procedure and the resulting evolution equations are presented in Appendix \ref{sec:perts}.   The annihilation terms in Eq.~(\ref{bkgd}) induce a coupling between $\delm$ and $\delr$, and these interactions are included in the perturbation equations.  Elastic scatterings between dark matter particles and relativistic particles lead to additional couplings between $\delm$ and $\delr$; these effects are considered in Section \ref{sec:cut} along with the suppression of perturbations due to the free-streaming of the dark matter particles. 

\begin{figure}
 \centering
 \resizebox{3.4in}{!}
 {
      \includegraphics{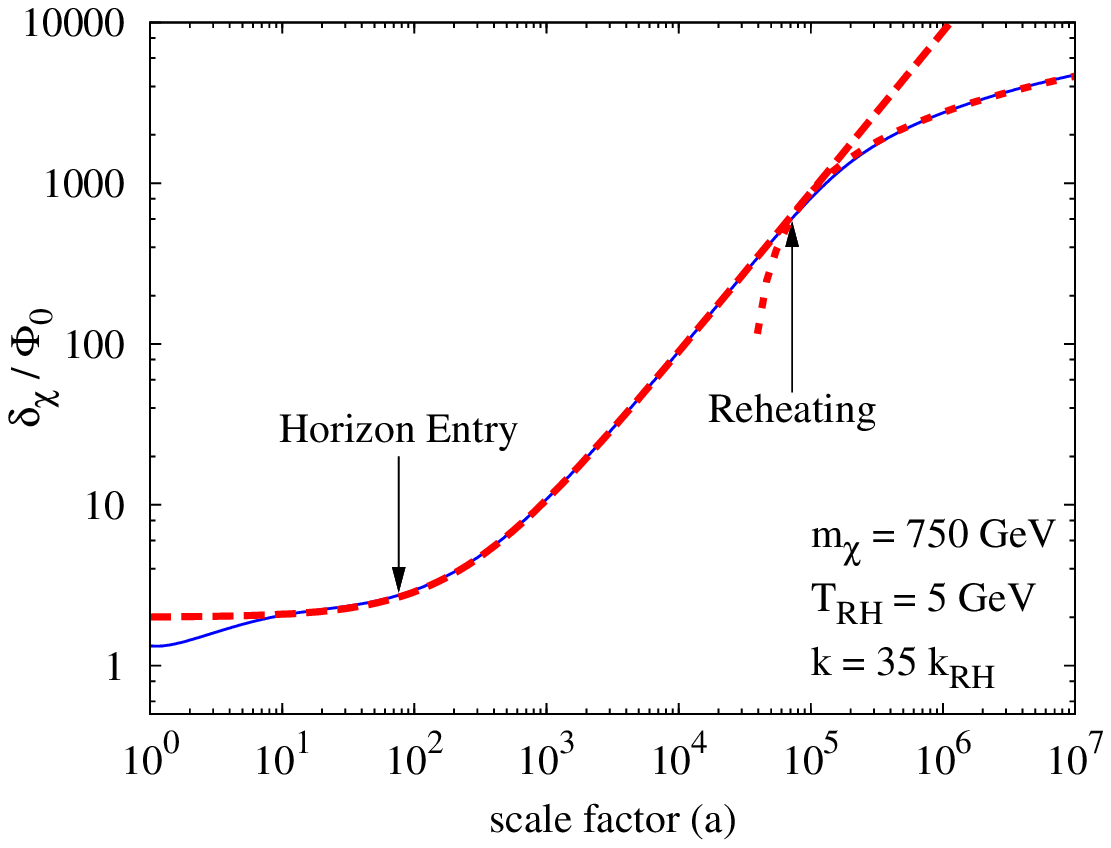}
 }
\caption{The evolution of the density perturbation in dark matter for a mode that enters the horizon prior to reheating ($k = 35 \kdec)$ in the freeze-in scenario ($\sigv = 2.9\times10^{-33} \sigunits$).  The background evolution for this scenario is shown in Figure \ref{bkgd} with $a_I = 3.6\times10^{-4}$.  The solid curve shows $\delm$.  The long-dashed curved shows the pre-reheating evolution of $\dels$, and the short-dashed curve shows the post-reheating evolution of $\delm$ predicted by Eq.~(\ref{delmloggrowth}) with $a_\mathrm{log} = 1.29\adec. $}
\label{Fig:DelMFI}
\end{figure}
\begin{figure*}
 \centering
\begin{minipage}{0.5\textwidth}
\centering
 \resizebox{3.4in}{!}
 {
      \includegraphics{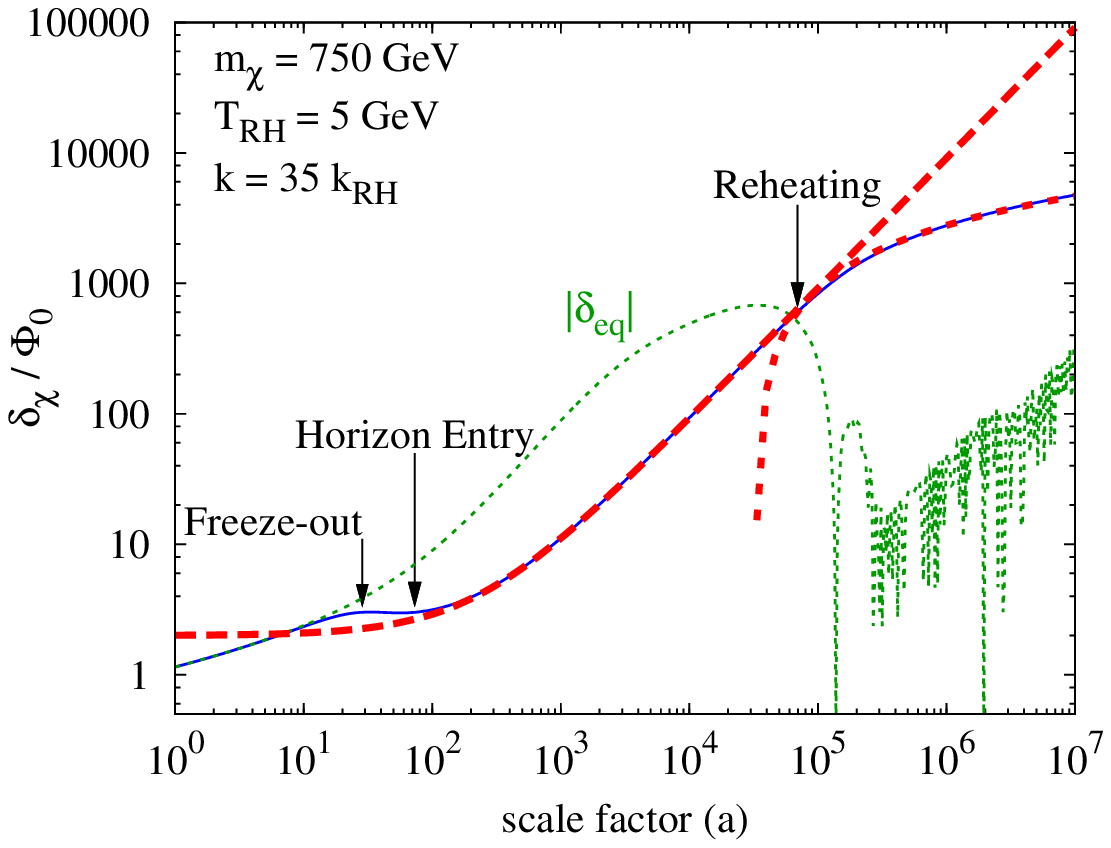}
 }
\end{minipage}%
\begin{minipage}{0.5\textwidth}
\centering
 \resizebox{3.4in}{!}
{
      \includegraphics{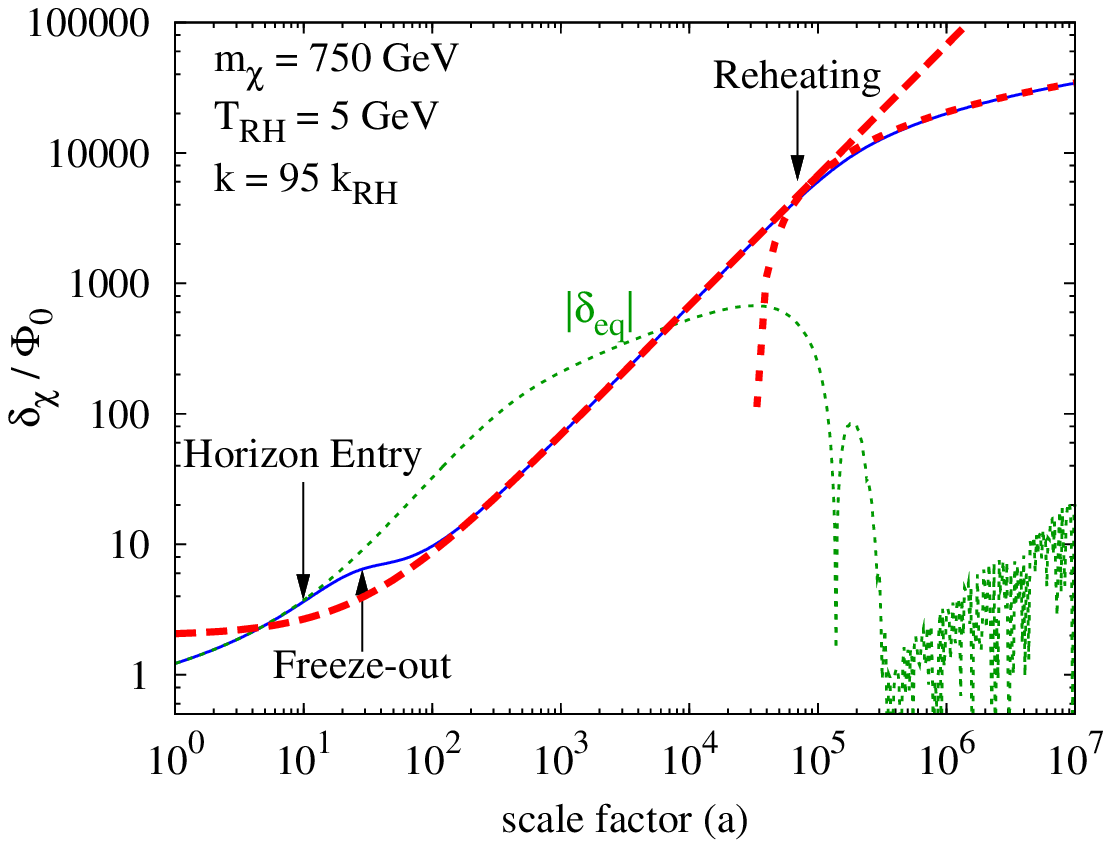}
 }
\end{minipage}%
\caption{The evolution of the density perturbation in dark matter for modes that enter the horizon prior to reheating ($k = 35 \kdec$ in the left panel and $k = 95 \kdec$ in the right panel) in the freeze-out scenario ($\sigv = 9.0\times10^{-31} \sigunits$).  The background evolution for this scenario is shown in Figure \ref{bkgd} with $a_I = 3.4\times10^{-4}$.  The solid curve shows $\delm$.  The long-dashed curved shows the pre-reheating evolution of $\dels$, and the short-dashed curve shows the post-reheating evolution of $\delm$ predicted by Eq.~(\ref{delmloggrowth}) with $a_\mathrm{log} = 1.29\adec.$ }
\label{Fig:DelMFO}
\end{figure*}

The perturbation evolution equations are solved numerically for each $k$ value starting before the mode enters the horizon and after the dark matter particle becomes nonrelativistic.  Since the scalar field is the only source of energy at the start of the EMDE, the perturbations are adiabatic and are all linearly related to the initial perturbation in the spacetime curvature $\Phi_0$: see Appendix \ref{sec:ics} for details.  Figure \ref{Fig:DelMFI} shows the evolution of the dark matter density perturbation $\delm$ for a mode with wave number $k = 35 \kdec$, where $\kdec \equiv \adec H(\adec)$ is the wave number of the mode that enters the horizon at reheating.  The background evolution for this cosmological scenario is shown in Figure \ref{Fig:bkgd}; the value of $a_I$ in this figure is determined by setting $a=1$ at the start of the numerical integration of the perturbation equations.  In Figure \ref{Fig:DelMFI}, $\sigv = 2.9\times10^{-33} \sigunits$, and the dark matter freezes in; the comoving number density of dark matter particles is conserved for $a \gsim 10$, which corresponds to $a/a_I \gsim 3 \times 10^4$ in Figure \ref{Fig:bkgd}.  When pair production halts, the mode shown in Figure \ref{Fig:DelMFI} has not yet entered the horizon.  Since both $\rhos$ and $\rhom$ are proportional to  $a^{-3}$ at this point onwards, adiabaticity demands that $\delm = \dels$ on superhorizon scales.  The long-dashed line in Figure \ref{Fig:DelMFI} shows the evolution of $\dels$ prior to reheating [see Eq.~(\ref{scalarics:dels}) in Appendix \ref{sec:ics}]; we see that $\delm \rightarrow \dels$ when $a \simeq 10$.

The evolution of the dark matter perturbation in the freeze-out scenario is shown in Figure \ref{Fig:DelMFO}.  While the dark matter is in thermal equilibrium, $\delm \simeq \deleq$, where
 $\deleq \equiv \nchieq/\nchieq^0 - 1$ is the perturbation in the equilibrium density of the dark matter.  Since $\nchieq$ is determined by the radiation density, $\deleq$ is related to the perturbation in the radiation density:
\beq
\deleq = \frac{1}{4} \left(\frac{3}{2} + \frac{m_\chi}{T}\right)\delr.
\label{deleq}
\eeq
The evolution of $\deleq$ is also shown in Figure \ref{Fig:DelMFO}.  While the mode is outside the horizon, $\delr$ is constant, but $\deleq$ increases as $\mdm/T$ increases.  After the mode enters the horizon, $\delr$ begins to grow as it is sourced by the increasingly inhomogeneous decaying scalar field.  However, the amplitude of $\delr$ plummets during reheating, and after reheating, $\delr$ oscillates with a constant amplitude that is far smaller than it would have in the absence of an EMDE.  This strange behavior of the radiation perturbation was discovered in Ref. \cite{ES11} and later confirmed in Refs. \cite{BR14, FOW14}.  Since $\deleq \simeq 0.25(\mdm/T)\delr$ while the dark matter is nonrelativistic, $\deleq$ also grows prior to reheating and then decreases rapidly before beginning oscillations, as seen in Figure \ref{Fig:DelMFO}.    If the dark matter freezes out (or kinetically decouples) after reheating, this reduction in $\delr$ would further suppress the small-scale dark matter density fluctuations.  However, it is doubtful that this additional suppression on top of the usual suppression on scales smaller than the decoupling scale would be observable.  Instead, I will focus on scenarios in which the dark matter freezes out and kinetically decouples prior to reheating.  

Figure~\ref{Fig:DelMFO} shows the evolution of two such perturbation modes, one that freezes out prior to horizon entry ($k = 35 \kdec$) and one that freezes out after horizon entry ($k = 95 \kdec$).  In both cases, $\delm \simeq \deleq$ until $H = \sigv \nchieq$, which is marked as ``Freeze-out" in Figure \ref{Fig:DelMFO}.  After freeze-out, $\delm$ matches the evolution of $\dels$, as it did in the freeze-in case shown in Figure \ref{Fig:DelMFI}.   In the absence of pair production and annihilations, the scalar perturbations and the dark matter perturbations are governed by the same set of equations prior to reheating: see Eq.~(\ref{perts}).  The solution to these equations is $\delta(a) = \delta_0+ (2/3)\tilk^2 \Phi_0 a$, where $\delta_0$ is a constant and $\tilk \equiv k/H(a=1)$.  Therefore, after freeze-out or freeze-in and before reheating, $\delm$ and $\dels$ differ by a constant, and that constant becomes less and less important as both $\delm$ and $\dels$ grow linearly with the scale factor.  By the time reheating occurs, $\delm \simeq \dels \simeq  (2/3)\tilk^2 \Phi_0 a$.  The same relation applies to perturbations in the density of nonthermal dark matter \cite{ES11}.  

When the Universe becomes radiation dominated, $\delm$ begins to grow logarithmically and $a \delm'(a)$ is constant.  If the transition from linear to logarithmic growth occurs when $a = a_\mathrm{log}$, then 
\beq
\delm(k>\krh) = \frac{2}{3} a_\mathrm{log} \tilk^2 \Phi_0 \left[1 + \ln\left(\frac{a}{a_\mathrm{log}}\right)\right]
\label{delmloggrowth}
\eeq
during radiation domination.  Figures \ref{Fig:DelMFI} and \ref{Fig:DelMFO} show that this model fits the numerical solution for $\delm$ when $a_\mathrm{log} = 1.29 \adec$, which is the same post-reheating behavior found for nonthermal dark matter density perturbations in Ref. \cite{ES11}.  When $a_\mathrm{log}$ is defined in this way, the numerical solution to Eq.~(\ref{bkgd}) reveals that \mbox{$[k_\mathrm{RH,h}/H(a=1)]^2 = 1/a_\mathrm{log}$}, so Eq.~(\ref{delmloggrowth}) may be rewritten as
\beq
\delm(k>\krh) = \frac{2}{3} \left(\frac{k}{\kdec}\right)^2 \Phi_0 \left[1 + \ln\left(\frac{a/\adec}{1.29}\right)\right].
\label{linmodel}
\eeq

\begin{figure}
 \centering
 \resizebox{3.4in}{!}
 {
      \includegraphics{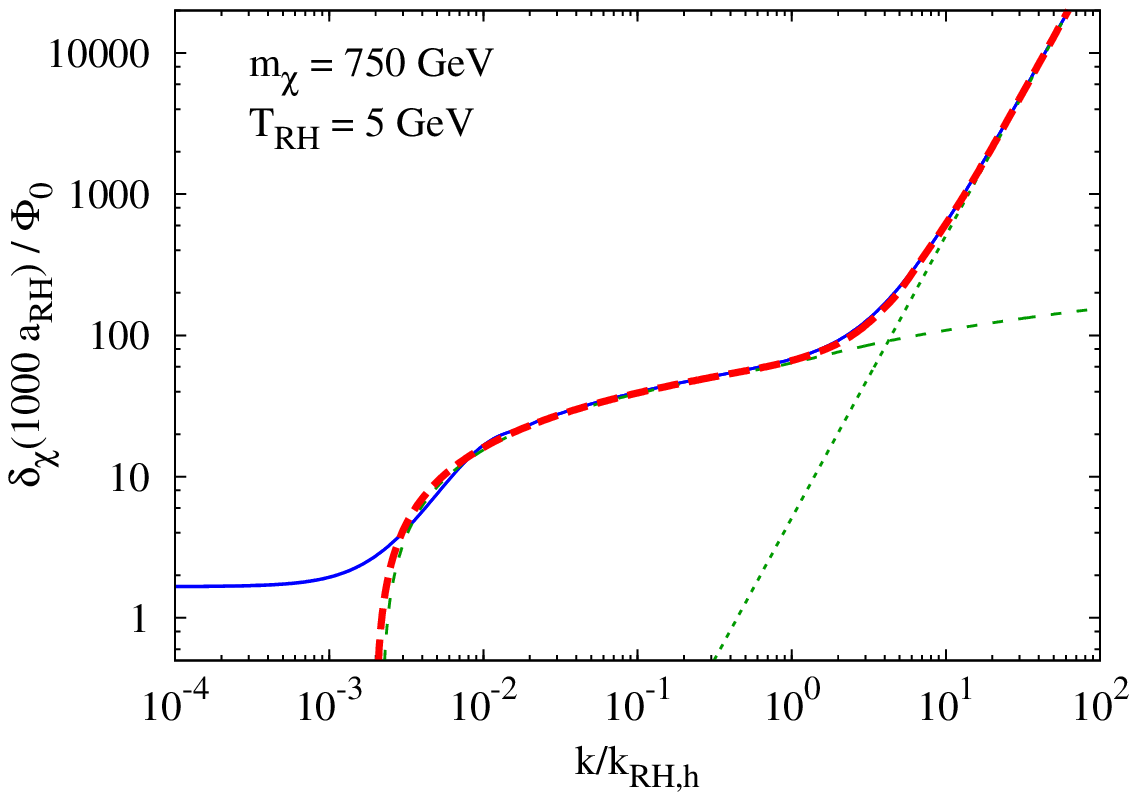}
 }
\caption{The dark matter density perturbation $\delm$ evaluated long after reheating as a function of the perturbation's wave number ($a = 1000\adec$); modes with $k/\kdec < 0.001$ have not yet entered the horizon.  The solid curves show the numerical solutions for both values of $\sigv$ shown in Figure \ref{Fig:bkgd}; there is no distinguishable difference between these two $\sigv$ values.  The thin dotted line shows Eq.~(\ref{linmodel}), while the thin dashed line shows Eq.~(\ref{logmodel}) with $A=9.11$ and $B=0.594$.  The thick dashed line shows Eq.~(\ref{logmodel}) with the fitting functions $A(k/\kdec)$ and $B(k/\kdec)$ derived for nonthermal dark matter perturbations in Ref. \cite{ES11}.}
\label{Fig:DeltaMk}
\end{figure}

Figure~\ref{Fig:DeltaMk} shows $\delm$ evaluated when $a = 1000 \adec$ as a function of the perturbation's wave number; we see that Eq.~(\ref{linmodel}) successfully predicts the post-reheating amplitude of $\delm$ on all scales that enter the horizon during the EMDE (i.e. modes with $k \gg \kdec$).  Figure~\ref{Fig:DeltaMk} also shows that the modes that have not yet entered the horizon have $\delm = 5/3 \Phi_0$, which corresponds to an adiabatic perturbation during radiation domination.  The subhorizon modes in Figure \ref{Fig:DeltaMk} that entered the horizon after reheating follow the standard evolution of density perturbations in a radiation-dominated Universe with $\delta\rho_r \gg \delta\rho_\chi$:
\beq
\delm(k<\krh) = \frac{10}{9} \Phi_0 \left[A \ln \left(B \frac{a}{a_\mathrm{hor}}\right)\right]
\label{logmodel}
\eeq
with $A=9.11$, $B=0.594$, and $k = a_\mathrm{hor}H(a_\mathrm{hor})$ \cite{HS96}. 

Figure~\ref{Fig:DeltaMk} indicates that the correspondence between perturbations in thermal and nonthermal dark matter extends to perturbations of all wave numbers, regardless of the value of $\sigv$.  Solving the perturbation equations for other values of $\sigv$ confirms that as long as freeze-out or freeze-in occurs well before reheating ($T_f \gsim 5 T_\mathrm{RH})$, the post-reheating evolution of $\delm$ matches Eq.~(\ref{linmodel}).  Therefore, the transfer function for density perturbations derived in Ref.~\cite{ES11} can be applied to perturbations in thermal dark matter.  This transfer function was derived by finding functions $A(k/\kdec)$ and $B(k/\kdec)$ that match the behavior of $\delm(k/\kdec)$ when inserted in Eq.~(\ref{logmodel}).  The subsequent evolution of the density perturbation was determined by matching Eq.~(\ref{logmodel}) to the growing and decaying modes of the Meszaros equation \cite{Meszaros74, HS96}.  It follows that the impact of an EMDE on the matter transfer function is to take $T(k)\rightarrow R(k)T(k)$, where
\beq
R(k) = \frac{A(k/\kdec) \ln \left[\left(\frac{4}{e^3}\right)^\frac{f_2}{f_1}\frac{B(k/\kdec) a_\mathrm{eq}}{a_\mathrm{hor}(k)}\right]}{9.11 \ln \left[\left(\frac{4}{e^3}\right)^\frac{f_2}{f_1} 0.594  \frac{\sqrt{2} k}{k_\mathrm{eq}}\right]} \\
\label{TRdef}
\eeq
for $k > 0.05 \kdec$ and $R=1$ for $k \leq 0.05 \kdec$.  In this expression, the subscript ``eq" refers to matter-radiation equality, $f_1$ and $f_2$ are constants determined by the baryon fraction, and a fitting function is used to find $a_\mathrm{hor}(k)$ for all modes: see Ref.~\cite{ES11} for details.  

Since the reheat temperature only enters the transfer function through the value of $\kdec$, it is useful to derive an expression that directly relates \mbox{$\kdec = \adec H(\adec)$} to $\Trh$.  Numerically solving Eq.~(\ref{bkgd}) for a variety of reheat temperatures reveals that there is a consistent relationship between $H(\adec)$ and the value of $H$ in a radiation-dominated universe with temperature $\Trh$: $H(\adec) = 0.88  \sqrt{8\pi G \rho_r(T_\mathrm{RH})/3}$.  Furthermore, $\rhor(a)a^{4} = 1.087 \rho_r(T_\mathrm{RH})\adec^4$ long after reheating, which quantifies the amount of entropy injected into the radiation bath after $a=\adec$.  It follows from the conservation of entropy that $g_{*S}(T) T^3 a^3 = 1.087^{3/4} g_{*S}(\Trh) \Trh^3 \adec^3$ for all $T\ll \Trh$, where $g_{*S}$ is the effective number of degrees of freedom that contribute to the entropy density $s$: $g_{*S}\equiv s/[(2 \pi^2/45)T^3]$.  These two results indicate that $\kdec = 0.86 \krh$, where $\krh$ is the wave number of the mode that enters the horizon when the temperature of a radiation-dominated universe equals $\Trh$.  If $T_0$ is the temperature of the radiation today, then 
\begin{align}
\krh &\equiv \left[\frac{g_{*S}(T_0)}{g_{*S}(\Trh)}\right]^{1/3} \frac{T_0}{\Trh} \sqrt{\frac{8\pi G}{3} \left(\frac{\pi^2}{30}\gstarRH \Trh^4\right);} \label{krh}\\
&=0.0117\left(\frac{T_\mathrm{RH}}{1 \, \mbox{MeV}}\right)\left(\frac{10.75}{g_{*S,\mathrm{RH}}}\right)^{1/3} \left(\frac{\gstarRH}{10.75}\right)^{1/2}  \, \mbox{pc}^{-1}. \nonumber
\end{align}
To compute the transfer function associated with a specific reheat temperature, $\kdec = 0.86 \krh (\Trh)$ should be used in Eq.~(\ref{TRdef}).\footnote{In Ref. \cite{ES11}, $\kdec = \krh$ was used to define $T_\mathrm{RH}$, which does not give the same reheat temperature as Eq.~(\ref{Trhdef}).  The reheat temperatures quoted in Ref. \cite{ES11} should be divided by a factor of 0.86 to obtain $\Trh$ as defined by Eq.~(\ref{Trhdef}).}  

\begin{figure}
 \centering
 \resizebox{3.4in}{!}
 {
      \includegraphics{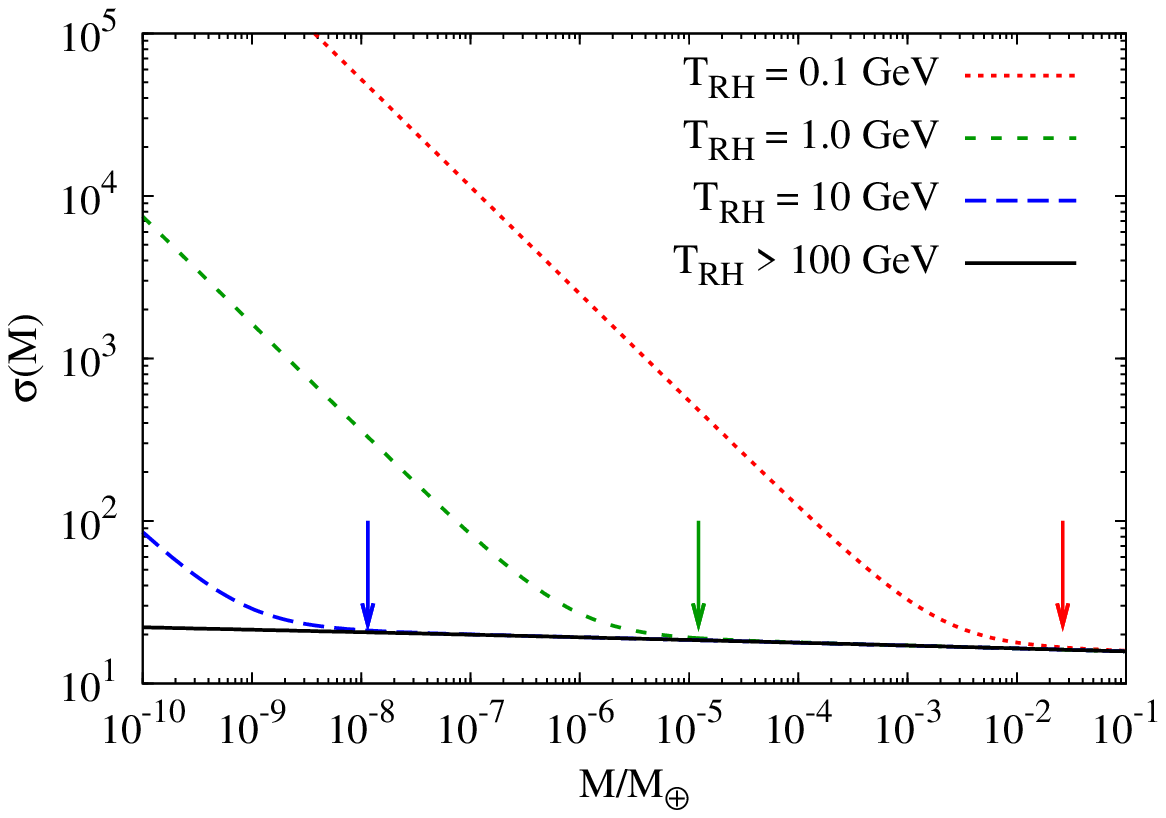}
 }
\caption{The impact of an EMDE on $\sigma(M)$: the present-day rms density perturbation in spheres containing mass $M$.  The arrows show $\mdec$, as defined by Eq.~(\ref{mrh}), for the three values of $\Trh$ shown.  In each case, the EMDE does not affect $\sigma(M)$ for $M>\mdec$, but it significantly enhances $\sigma(M)$ for smaller masses.}
\label{Fig:SigM}
\end{figure}

The scales that are affected by an EMDE are smaller than the baryon Jeans length.  Since baryons do not participate in gravitational collapse on these scales, these perturbations grow at a slower rate than large-scale perturbations \cite{HS96}.  Ref.~\cite{ES11} derived a scale-dependent growth function $D(k,z)$ that accounts for this suppression of the growth of small-scale perturbations.  This growth function, combined with the modified matter transfer function $T(k)$, is used to calculate the rms density perturbation in a sphere that contains a mass $M$:
\beq
\sigma^2(M,z) = \int \frac{\drm^3k}{(2\pi)^3} \left[D(k,z)T(k)\right]^2 P_\mathrm{p}(k) F^2(kR).
\label{sigmadef}
\eeq
In this expression, $P_\mathrm{p}(k)$ is the power spectrum of superhorizon density perturbations during radiation domination, and \mbox{$R = [3M/(4\pi \rho_{\mathrm{m},0})]^{1/3}$}, where $\rho_{\mathrm{m},0}$ is the present-day matter density.   The window function $F(kR)$ is a convolution of a top-hat window function and a Gaussian window function,
\beq
F(kR) = \exp\left[-\frac{k^2(\alpha R)^2}{2}\right]
\times\frac{3\left[\sin(kR)-(kR)\cos(kR)\right]}{(kR)^3},
\eeq
with $\alpha = 0.0001$.  As discussed in Ref.~\cite{ES11}, the inclusion of this Gaussian window function prevents modes with $k^{-1}\ll R$ from contributing to $\sigma^2(M)$ while leaving $\sigma^2(M)$ unaltered for $M>\mdec$.
To facilitate comparison with previous work, $\sigma(M)$ is computed using the same WMAP7 \cite{wmap7param} cosmological parameters used in Ref. \cite{ES11}: $\Omega_m h^2 =0.135$, $h=0.704$ and \mbox{$\Delta_{\cal R}^2(k) = (2.44\times10^{-9})[k/(0.002\, \mathrm{Mpc}^{-1})]^{0.963-1}$}.  The slight difference between these parameters and the parameters determined by the \emph{Planck} mission \cite{2015planck} would have a negligible impact on the results of this analysis and do not justify altering the transfer function.

Figure \ref{Fig:SigM} shows the present-day value of $\sigma(M)$ for different reheat temperatures.  An EMDE alters $\sigma(M)$ for masses that are smaller than 
\begin{align}
\mdec &\equiv \frac{4\pi}{3} \rho_{\mathrm{m},0} \krh^{-3}\label{mrh} \\
 &= 32.7 M_\oplus\! \left(\frac{10 \, \mbox{MeV}}{T_\mathrm{RH}}\right)^3\!\left(\frac{g_{*S}[T_\mathrm{RH}]}{10.75}\right)\!\!\left(\frac{10.75}{g_*[T_\mathrm{RH}]}\right)^{3/2}\!\!\!\!\!\!\!.\nonumber
\end{align}
Ref.~\cite{ES11} used the Press-Schechter mass function \cite{PS74} to predict how this enhancement of $\sigma(M)$ at small masses would affect the microhalo population.  They found that an EMDE dramatically enhances the number of microhalos with masses less than $\mdec$.  Moreover, these microhalos form very early; at a redshift $z=100$, over half of the dark matter is bound into microhalos with masses greater than 0.001 $\mdec$.  The fraction of dark matter bound into halos is equally large at higher redshifts, but the microhalos are smaller.  The size of the first and smallest microhalos is determined by the matter power spectrum, which usually has an exponential cut-off on small scales: $\delm(k) \propto \exp[-k^2/(2\kcut^2)]$.  For nonthermal dark matter, the cut-off scale $\kcut$ is determined by the average velocity of a dark matter particle at reheating and is therefore related to the mass difference between the scalar and dark matter \cite{ES11}.  For thermal dark matter, elastic scatterings between dark matter particles and relativistic leptons keep the dark matter and radiation in kinetic equilibrium after the dark matter thermally decouples.  As shown in the next section, $\kcut$ is determined by these interactions.

\section{Kinetic Decoupling and Free-Streaming}
\label{sec:cut}

The momentum transfer rate for dark matter particles elastically scattering off relativistic particles with temperature $T$ and number density $n_\mathrm{rel}$ is 
\beq
\Gamma_\mathrm{el} = \langle \sigma_\mathrm{el}v \rangle n_\mathrm{rel} \frac{T}{\mdm};
\eeq
the $T/\mdm$ factor accounts for the fact that it takes $\mdm/T$ elastic scatterings to significantly alter the dark matter particle's momentum.  As long as $\Gamma_\mathrm{el} > H$, the dark matter will remain in kinetic equilibrium with the radiation bath, and the perturbations in the dark matter density will be coupled to the perturbations in the radiation density.  Numerous studies have considered how dark matter decouples from radiation during a period of radiation domination \cite{GHS04, LZ05, PSK2006, Bertschinger06, BH07}: since the radiation perturbations experience damped oscillations in a radiation-dominated universe, dark matter perturbations on scales that enter the horizon prior to decoupling are suppressed.  After decoupling, the residual velocities of the dark matter particles lead to a free-streaming effect that further suppresses the amplitude of perturbations on scales smaller than the free-streaming distance.  Both suppression scales are directly related to the temperature at which $H=\Gamma_\mathrm{el}$.  In a radiation-dominated universe, this decoupling temperature is defined by the relation 
\beq  
\Gamma_\mathrm{el} (\Tkds) \equiv \sqrt{\frac{8\pi G}{3} \left[\frac{\pi^2}{30}g_*(\Tkds) \Tkds^4\right]}.
\label{TkdSdef}
\eeq
The subscript $S$ denotes that $\Tkds$ is the temperature at which the dark matter particle would decouple in a standard thermal history; as shown below, the actual decoupling temperature $\Tkd$ greatly exceeds $\Tkds$ if dark matter decouples during an EMDE.  However, since $\Tkds$ has been calculated for many dark matter candidates, it provides a convenient way to parametrize the microphysics that determines the elastic scattering rate.

If $\Tkds < \Trh$, then the dark matter remains coupled to the radiation throughout reheating.  As previously mentioned, the perturbations in the radiation decrease rapidly when the Universe becomes radiation dominated, and if the dark matter is coupled to radiation at that time, its perturbations would most likely be similarly suppressed.  However, if $\Tkds > \Trh$, then the dark matter decouples during the EMDE.  When the radiation is at a certain temperature during the EMDE, the added density of the scalar field makes the expansion rate faster than it would be in a radiation-dominated universe with the same temperature.  Therefore, $H(\Tkds) > \Gamma_\mathrm{el}(\Tkds)$ if $\Tkds > \Trh$, and the dark matter must decouple at a higher temperature. 

The dependence of the kinetic decoupling temperature on the reheat temperature was first noted in Ref. \cite{GG08}; if $\sigma_\mathrm{el} \propto (T/\mdm)^2$, as is the case for most neutralinos, they concluded that dark matter decouples when the radiation temperature is $\Tkd \simeq \Tkds^2/\Trh$ if $\Tkds>\Trh$.  This decoupling temperature follows from equating $\Gamma_\mathrm{el}$ to the expansion rate during the EMDE.  The definition of $\adec$ implies that
\beq
H(a) =\left(\frac{a}{\adec}\right)^{-3/2} \sqrt{\frac{8\pi G}{3} \left(\frac{\pi^2}{30}\gstarRH \Trh^4\right)}.
\label{Hemde}
\eeq
For $a \gg a_I$, Eq.~(\ref{rhorad}) provides an expression for the temperature of the radiation bath during the EMDE as a function of the scale factor,
\beq
\frac{T}{\Trh} = \left[\frac25 \frac{\gstarRH}{g_*(T)}\right]^{1/4} \left(\frac{a}{\adec}\right)^{-3/8},
\label{temp}
\eeq
which can be used to obtain $H(T)$ during the EMDE:
\beq
H(T) = \frac52 \frac{g_*(T)T^4}{\gstarRH \Trh^4} \sqrt{\frac{8\pi G}{3} \left(\frac{\pi^2}{30}\gstarRH \Trh^4\right)}.
\eeq
Although $H\propto T^4$ during the EMDE, this equation shows that $H(T)$ does not equal $(T/\Trh)^4 \sqrt{8\pi G \rhor(\Trh)/3}$; this commonly used relation is not correct because the radiation temperature when $a=\adec$ does not equal $\Trh$, as can be seen in Eq.~(\ref{temp}).  If $\sigma_\mathrm{el} \propto (T/\mdm)^2$, then $\Gamma_\mathrm{el} \propto T^6$, and Eq.~(\ref{TkdSdef}) implies that $\Gamma_\mathrm{el}(\Tkd) = H(\Tkd)$ during the EMDE if
\beq
\Tkd = \left[\frac{g_*(\Tkd)^2}{g_*(\Tkds)\gstarRH}\right]^{1/4}\sqrt{\frac52} \frac{\Tkds^2}{\Trh}.
\label{Tkd}
\eeq
Since $\Tkd > \Tkds > \Trh$ and $g_*(T)$ monotonically decreases as the Universe cools, we see that including changes in the number of relativistic degrees of freedom and the exact function of $H(T)$ during an EMDE implies that the dark matter decouples at an even higher temperature than predicted in Ref. \cite{GG08}.

\begin{figure}
\centering
\resizebox{3.4in}{!}
 {
      \includegraphics{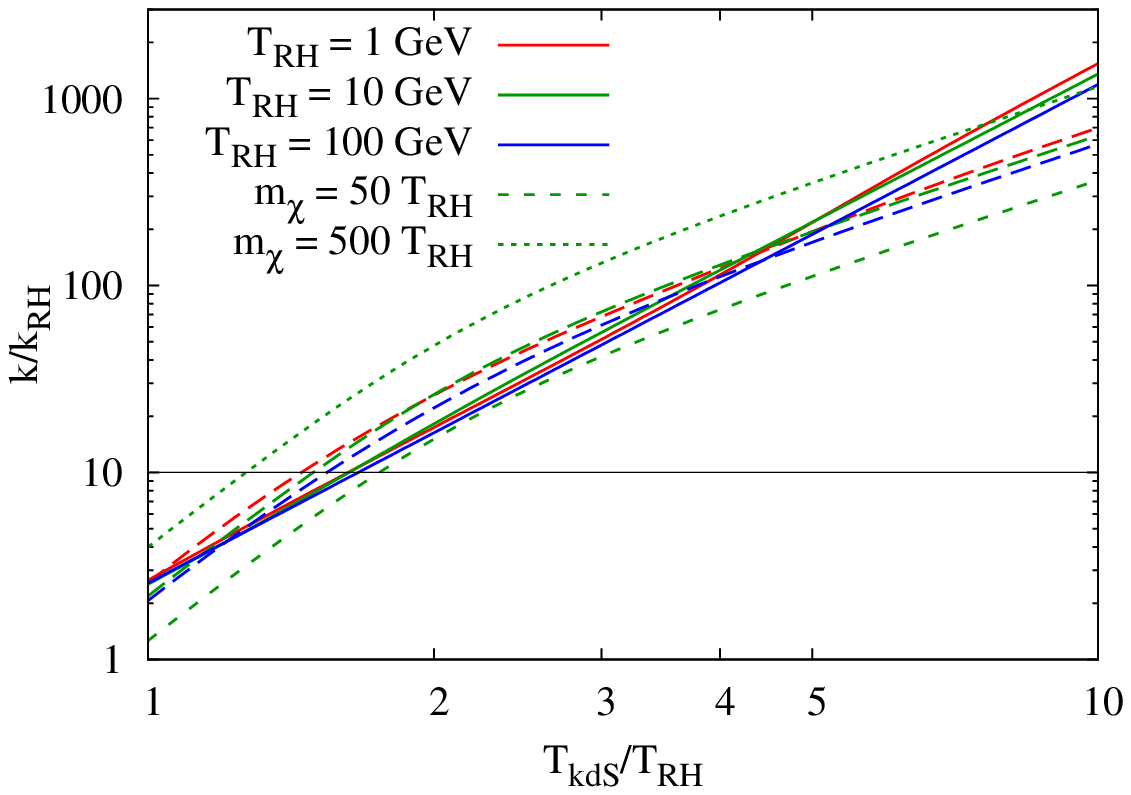}
 }
\caption{The kinetic decoupling and free-streaming scales as a function of $\Tkds/\Trh$, where $\Tkds$ is the kinetic decoupling temperature in a radiation-dominated universe.  The solid curves show $\kkd/\krh$, where $\kkd$ is the wave number of the mode that enters the horizon when the expansion rate equals the elastic scattering rate.  The long-dashed curves show $\kfs/\krh$, where $\kfs$ is the inverse of the free-streaming horizon, for the same reheat temperatures as the solid lines of the same color and $\mdm = 150 \Trh$.  The short-dashed and dotted lines show how $\kfs/\krh$ changes if the dark matter particle mass changes while $\Trh$ is held fixed at 10 GeV.}
\label{Fig:Cut}
\end{figure}

Elastic scatterings between dark matter and relativistic particles affect perturbation modes that enter the horizon prior to decoupling, so it is useful to define $\kkd$ to be the wave number of the mode that enters the horizon when $T=\Tkd$:
\begin{align}
\frac{\kkd}{\krh} &\equiv \frac{a(\Tkd) H(\Tkd)}{\adec \sqrt{8\pi G \rhor(\Trh)/3}},\nonumber \\
&= \frac52 \frac{g_*(\Tkd)}{g_*^{1/3}(\Tkds){\gstarRH^{2/3}}} \left(\frac{\Tkds}{\Trh}\right)^{8/3}.
\label{kdrat}
\end{align} 
Figure~\ref{Fig:Cut} shows $\kkd/\krh$ as a function of $\Tkds/\Trh$ for $\Trh = 1$ GeV, 10 GeV, and 100 GeV.  This ratio is rather insensitive to $\Trh$ at these temperatures; since $g_*(T)$ varies by less than 30\% above temperatures of 1 GeV, $\kkd/\krh \simeq (5/2)(\Tkds/\Trh)^{8/3}$ when $\Trh > 1$ GeV.  At lower reheat temperatures, however, the $g_*$ factors in Eq.~(\ref{kdrat}) can increase $\kkd/\krh$ by up to a factor of 10, so the $\kkd/\krh$ curves shown in Figure \ref{Fig:Cut} are conservative.  Thus we see that $\kkd \gg \krh$ even if the standard decoupling temperature $\Tkds$ is only slightly higher than the reheat temperature.

After the dark matter decouples from the radiation bath, the velocities of the dark matter particles decrease as $1/a$.  Since $T \propto a^{-3/8}$ during the EMDE, the velocity of the dark matter decreases by a greater amount as the Universe cools if it decouples during the EMDE than if it decouples in a radiation-dominated universe.  The fact that the dark matter decouples earlier during an EMDE than it would in a radiation-dominated universe further reduces the final velocities of the dark matter particles.  When the Universe is radiation dominated and $T=\Trh$, the velocity of the dark matter particles is reduced by a factor of $(\Trh/\Tkds)^{23/6}$ from what it would have been if the dark matter particles decoupled when $T=\Tkds$ in a radiation-dominated universe.  Since the comoving free-streaming horizon is directly related to the velocity of the dark matter particles \cite{GHS05, Bertschinger06, BLRV09},
\beq
\lambda_\mathrm{fs} = \int^{t_0}_{t_\mathrm{kd}} \frac{v}{a} \drm t \simeq \sqrt{\frac{\Tkd}{\mdm}} a(\Tkd) \int_{a(\Tkd)}^{1} \frac{\drm a}{a^3 H(a)},
\label{lambdafs}
\eeq
dark matter particles that decouple prior to reheating have much smaller free-streaming horizons than they would if they had decoupled in a radiation-dominated universe \cite{GG08}. 

The free-streaming horizon is evaluated by breaking the integral in Eq.~(\ref{lambdafs}) into three parts, treating the transitions between the EMDE, radiation domination, and matter domination as instantaneous.  For $a<\adec$, Eq.~(\ref{Hemde}) provides $H(a)$.  After reheating, $\rhor(a)a^{4} = 1.087 \rho_r(T_\mathrm{RH})\adec^4$ [see discussion above Eq.~(\ref{krh})], so $H(a) = 1.04 (a/\adec)^{-2} \sqrt{8\pi G \rhor(\Trh)/3}$ until matter-radiation equality, after which $H(a) \propto a^{-3/2}$.  With these approximations,
\begin{align}
\lambda_\mathrm{fs} =& \sqrt{\frac{\Tkd}{\mdm}}\left(\frac{a_\mathrm{kd}}{\adec}\right) \frac{1}{\krh}  \\
&\times\left[2\left(\sqrt{\frac{\adec}{a_\mathrm{kd}}} - 1 \right) + \frac{\ln \left({a_\mathrm{eq}}/{\adec}\right)}{1.04} + \frac{2-2\sqrt{a_\mathrm{eq}}}{1.04}\right], \nonumber
\end{align}
where $a_\mathrm{kd} = a(\Tkd)$ and $a_\mathrm{eq} = 4.15\times10^{-5}/(\Omega_m h^2)$ is the scale factor at matter-radiation equality.  Free-streaming suppresses perturbations with wave numbers greater than $\kfs \equiv \lambda_\mathrm{fs}^{-1}$.  Using Eq.~(\ref{temp}) and Eq.~(\ref{Tkd}) to evaluate ${\adec}/{a_\mathrm{kd}}$ and $\Tkd$ yields
\begin{align}
\frac{\kfs}{\krh} =& \sqrt{\frac{\mdm}{\Tkds}}\left(\frac{\Tkds}{\Trh}\right)^{29/6} r^{13/6} \left[\frac52 \frac{g_*(\Tkd)}{\gstarRH}\right]^{2/3} \\
&\times\left[{2\left\{\left(\frac{\Tkds}{\Trh}\right)^{8/3}\left[r^4\frac52\frac{g_*(\Tkd)}{\gstarRH}\right]^{1/3} - 1 \right\} }\right. \nonumber\\
&\,\,\,\,\,\,\,\,\, \left.{+ \frac{1}{1.04} \ln \left(\frac{\gstarRH^{1/3}\Trh}{g_{*S,\mathrm{eq}}^{1/3} T_\mathrm{eq}}\right)+ 1.88}\right]^{-1},\nonumber
\end{align}
where $r \equiv \Tkd/(\Tkds^2/\Trh)$, which can be obtained from Eq.~(\ref{Tkd}).  Figure \ref{Fig:Cut} shows how $\kfs/\krh$ depends on $\mdm/\Trh$ and $\Tkds/\Trh$.  The rapid cooling of the dark matter particles during the EMDE makes $\kfs/\krh \gg 1$ even if $\Tkds$ is just slightly larger than $\Trh$.

To account for the effects of free-streaming and kinetic decoupling, it is customary to exponentially suppress perturbations with $k > \kcut$, where $\kcut$ is the smaller of $\kkd$ and $\kfs$ \cite{GHS04, LZ05, PSK2006, Bertschinger06, BH07}.  While free-streaming will always suppress perturbations, it is less clear that momentum exchange between the dark matter particles and relativistic particles will have the same effect during an EMDE.  As discussed in Section \ref{sec:pertevol}, radiation perturbations on subhorizon scales grow prior to reheating, so it is reasonable to expect that perturbations in the dark matter will experience similar growth prior to decoupling.   As long as the dark matter decouples prior to reheating, the growth of the matter perturbation during the EMDE will lead to an enhancement of the small-scale matter power spectrum.  Therefore, although modes with $k > \kkd > \krh$ may not follow Eq.~(\ref{linmodel}), they will not be exponentially suppressed, and a more extensive analysis of kinetic decoupling prior to reheating is required to determine their properties.   However, Figure \ref{Fig:Cut} shows that $\kfs \simeq \kkd$ for $\mdm \lsim 500 \Trh$, so modes with $k>\kkd$ are suppressed by free-streaming in most cases of interest.  Therefore, I will include the effects of both free-streaming and kinetic decoupling by multiplying the matter transfer function by $\exp[-k^2/(2\kcut^2)]$, with the understanding that setting $\kcut$ equal to $\kkd$ when $\kkd < \kfs$ may underestimate the matter density perturbations.

\begin{figure}
 \centering
 \resizebox{3.4in}{!}
 {
      \includegraphics{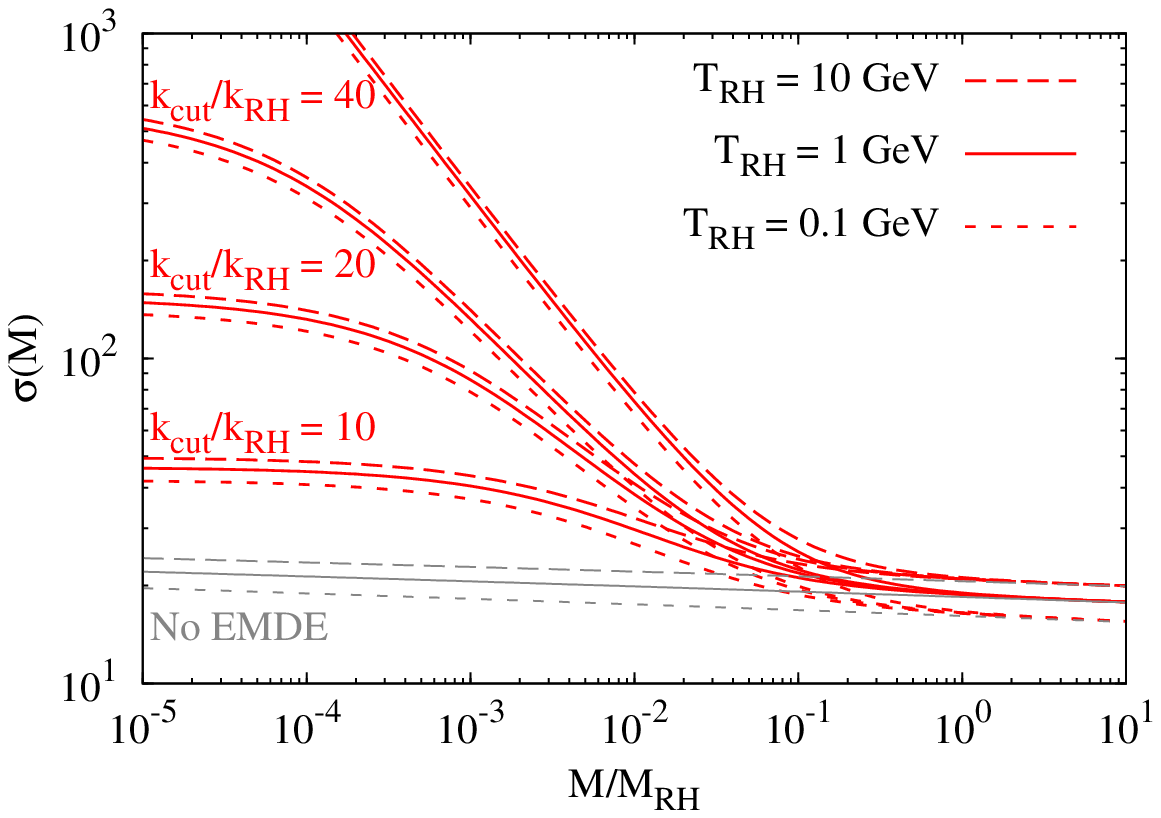}
 }
\caption{The rms density perturbation $\sigma(M)$ with an exponential cut-off in the matter power spectrum: \mbox{$\delta_m(k) \propto \exp[-k^2/(2\kcut^2)]$}.  The top set of curves have no cut-off, whereas the next three sets of curves have $\kcut/\krh$ = 40, 20, and 10 as marked.  The bottom set of curves shows $\sigma(M)$ in the absence of an EMDE with no cut-off; in this case $\Trh$ only determines $\mdec$.}
\label{Fig:SigSim}
\end{figure}

Figure \ref{Fig:SigSim} shows how a small-scale cut-off in the matter power spectrum affects the rms density perturbation $\sigma(M)$ for multiple values of $\Trh$ and $\kcut/\krh$.  Inserting a small-scale cut-off forces $\sigma(M)$ to level off for small masses, but it does not erase the effect of the EMDE if $\kcut/\krh \gsim 10$.  
Figure \ref{Fig:SigSim} also shows that $\sigma(M/\mdec)$ is not very sensitive to $\Trh$.  As seen in Figure \ref{Fig:SigM}, $\sigma(M)$ is a very shallow function of $M$ for $M > \mdec$, so changing $\mdec$ has only a small effect on $\sigma(M/\mdec)$.   In Section \ref{sec:boost}, I will show that the annihilation boost factor generated by an EMDE depends only on $\sigma(M/\mdec)$ and is therefore rather insensitive to $\Trh$.  The ratio $\kcut/\krh$ will play a much more significant role in the calculation of the boost factor, however.  Since both $\kfs/\krh$ and $\kkd/\krh$ are determined by $\Tkds/\Trh$, it is this ratio that will have the largest effect on the annihilation rate.  

\begin{figure}
 \centering
 \resizebox{3.4in}{!}
 {
      \includegraphics{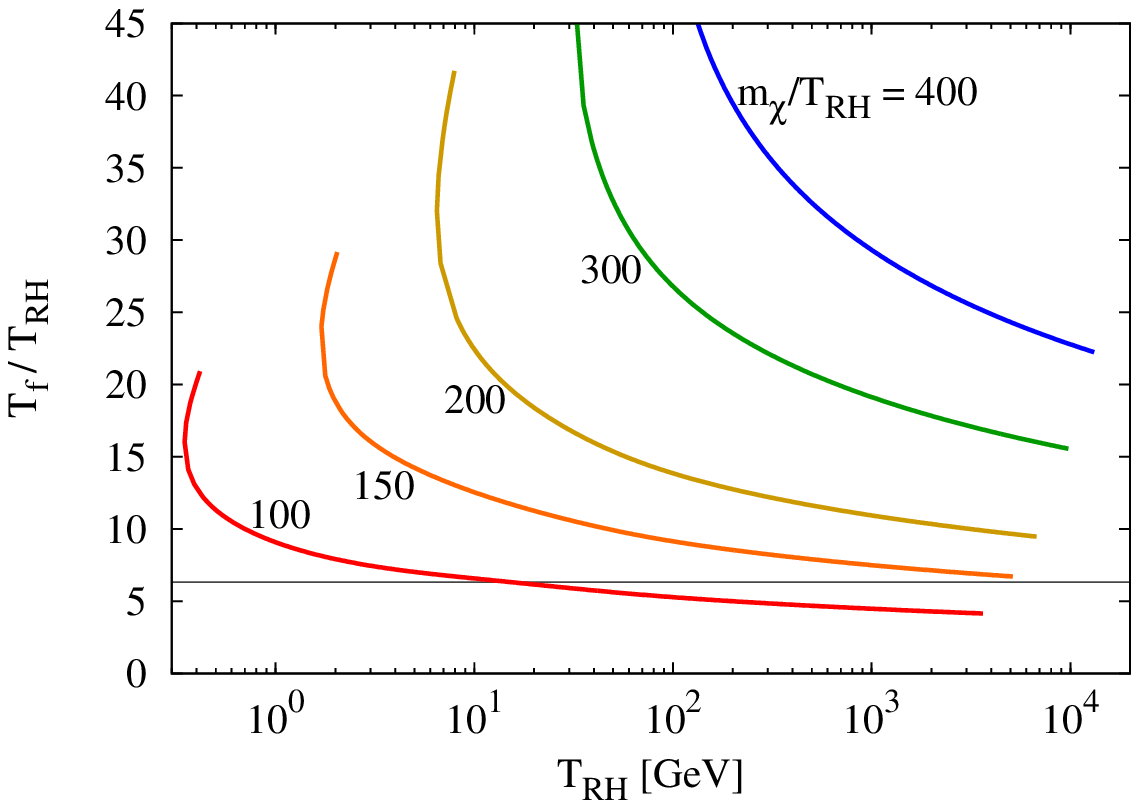}
 }
\caption{The ratio of the freeze-out temperature $T_f$ to the reheat temperature $\Trh$ for freeze-out scenarios that generate the observed dark matter abundance.  For each value of $\mdm/\Trh$ and $\Trh$, $\sigv$ was chosen so that Eq.~(\ref{OmegaFO}) gives $\Omega_\chi h^2 = 0.12$.  The horizontal line shows $T_f/\Trh = 2\sqrt{10}$, which is the minimum value of $T_f/\Trh$ required for the particles to kinetically decouple after freeze-out with $\kcut/\krh \gsim 10$.  The low-$\Trh$ end point of each curve is set by the minimum value of $\Trh$ required to generate $\Omega_\chi h^2 = 0.12$, while the high-$\Trh$ end point comes from demanding that $\sigv < \mdm^{-2}$.}
\label{Fig:Tfreeze}
\end{figure}

Figure \ref{Fig:Cut} shows that $\kcut/\krh \gsim 10$ if \mbox{$\Tkds/\Trh \gsim 2$}, which establishes a minimum value of \mbox{$\Tkds/\Trh$} required for an EMDE to enhance small-scale perturbations.  If $\Tkds/\Trh > 2$, then Eq.~(\ref{Tkd}) implies that $\Tkd/\Trh \gsim 2\sqrt{10}$.  Since dark matter kinetically decouples after it thermally decouples in most cases, freeze-out scenarios are expected to generate enhanced small-scale perturbations only if $T_f/\Trh > 2 \sqrt{10}$.  Figure \ref{Fig:Tfreeze} shows $T_f/\Trh$ for freeze-out scenarios that generate a relic abundance of dark matter that matches observations and still obey the unitarity condition $\sigv < \mdm^{-2}$.  Since dark matter particles freeze out when $\mdm \sim T/10$, $T_f / \Trh \sim 0.1\mdm/\Trh$ in most cases.  If $\mdm/\Trh$ is held fixed, there is still a mild dependence on $\Trh$ because as $\Trh$ increases, a slightly larger value of $\sigv$ is required to generate the observed dark matter abundance (see Figure \ref{Fig:OmegaDM}), which leads to a slightly lower $T_f/\mdm$ ratio.  Figure \ref{Fig:Tfreeze} shows that for $\mdm = 100 \Trh$, it is only possible to have $\Tkds/\Trh > 2$ and $\Tkd < T_f$ if the reheat temperature is less than 10 GeV, but $\Tkds/\Trh > 2$ and $\Tkd < T_f$ is possible for all freeze-out scenarios with $\mdm/\Trh >150$ that generate the observed dark matter abundance.  

Of course, the mere existence of $\Tkds$ values that satisfy both $\Tkds/\Trh > 2$ and $\Tkd < T_f$ does not imply that there are well-motivated particles that have the cross sections for annihilation and elastic scattering required to realize this scenario.  Indeed, most scans of WIMP parameter space predict $\Tkds$ values below a few GeV \cite{PSK2006}, but these scans are limited to models that generate the observed abundance of dark matter in a radiation-dominated universe.  A much smaller value of the annihilation cross section is required to thermally produce the observed abundance during an EMDE, and such particles would likely have weaker elastic interactions as well.  A search of SUSY parameter space for possible particles is left for future work.  In this analysis, I will restrict myself to scenarios with $T_f / \Trh > 2\sqrt{10}$ and use the ratio $\kcut/\krh$ to parametrize the impact of kinetic decoupling and free-streaming on the dark matter perturbations, with Figure \ref{Fig:Cut} serving as a guide for how to convert this ratio to a value of $\Tkds$.

\begin{figure}
 \resizebox{3.4in}{!}
 {
      \includegraphics{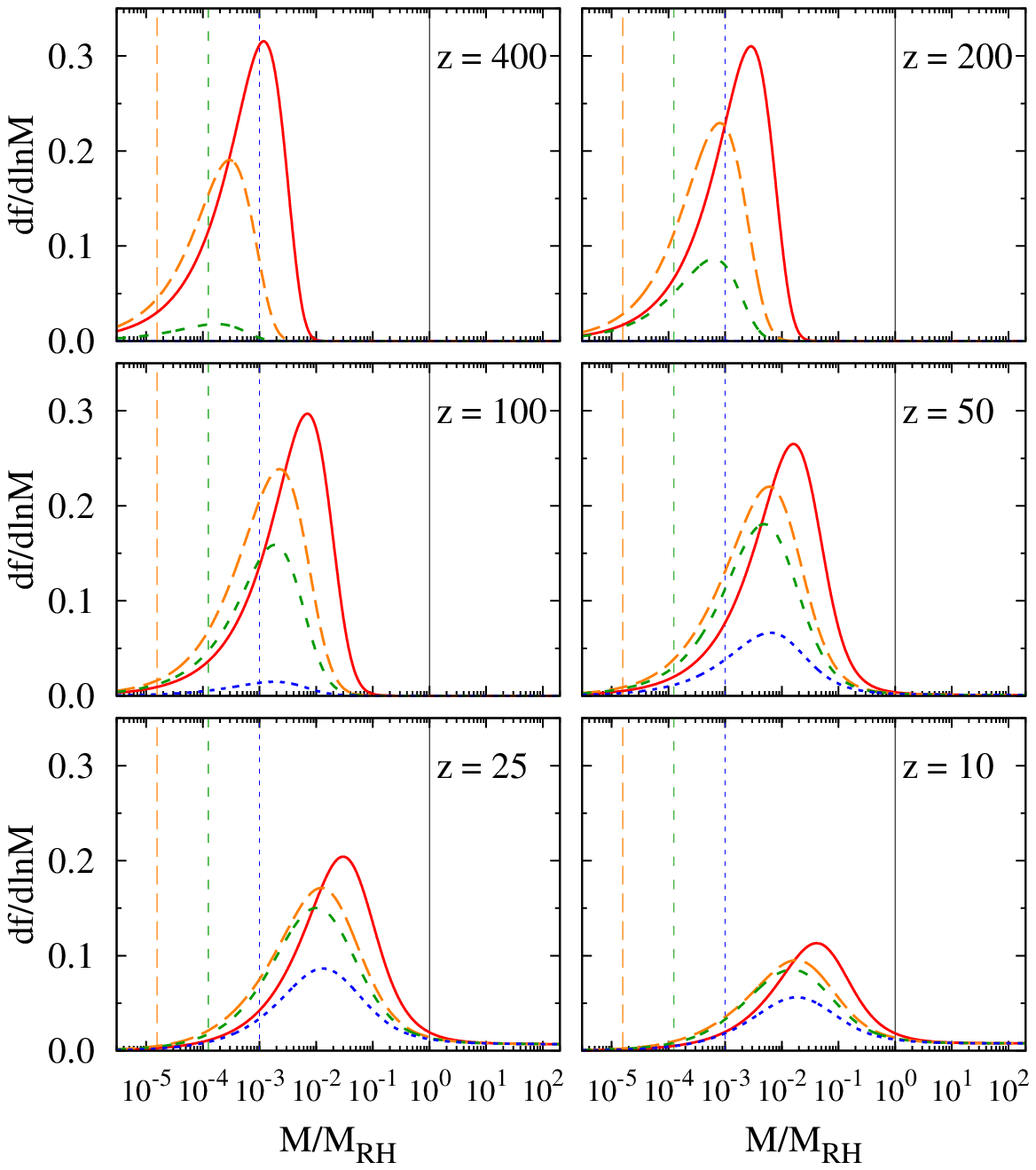}
 }
\caption{The evolution of the differential bound fraction given by Eq.~(\ref{dfdlnM}) for $\Trh = 5$ GeV.  At each redshift, the solid curve has no small-scale cut-off in the matter power spectrum, while the long-dashed, short-dashed, and dotted curves have $\kcut/\krh =$ 40, 20, and 10, respectively.  The solid vertical lines mark $M=\mdec$, and the other vertical lines show the cut-off masses $M_\mathrm{cut} = (4\pi/3) \rho_\mathrm{m,0} \kcut^{-3}$ for the curve with the same line type.}
\label{Fig:BMFevol}
\end{figure}

In the absence of a small-scale cut-off in the matter power spectrum, small scales will be increasingly homogeneous.  The Press-Schechter formalism \cite{PS74} predicts that microhalos form when $\sigma(M)$ exceeds the critical linear over-density $\delcoll$ ($\delcoll = 1.686$ for $z\gsim 2$).  Consequently, if $\sigma(M)$ increases without bound as $M$ decreases, microhalos will form at arbitrarily high redshift.  By limiting the amplitude of $\sigma(M)$ on small scales, the $\kcut/\krh$ ratio not only determines the mass of the smallest microhalos, it also determines when these microhalos will form.  Figure \ref{Fig:BMFevol} illustrates how $\kcut/\krh$ affects the evolution of microhalo population, which is characterized by the Press-Schechter differential bound fraction:
\beq
\frac{d f}{d \ln M} = \sqrt{\frac{2}{\pi}} \left|{\frac{d \ln\sigma}{d \ln M}}\right| \frac{\delcoll}{\sigma(M,z)} \exp\left[-\frac{\delcoll^2}{2\sigma^2(M,z)}\right].
\label{dfdlnM}
\eeq
The bound fraction \mbox{$f(M_1, M_2) \equiv \int_{\ln M_1}^{\ln M_2} \frac{df}{d\ln M} \, d\ln M$} gives the fraction of mass that is bound in halos with masses between $M_1$ and $M_2$.  Figure \ref{Fig:BMFevol} shows that most of the matter is already bound into microhalos at a redshift of 400 if $\kcut/\krh \gsim 40$.  For smaller values of $\kcut/\krh$, however, microhalo formation is postponed to later times: $z\simeq200$ for $\kcut = 20 \krh$ and $z\simeq50$ for $\kcut = 10 \krh$.   Decreasing the $\kcut/\krh$ ratio only slightly decreases the mass range of microhalos that are present at a given redshift, but if the masses of these microhalos are smaller than $M_\mathrm{cut} = (4\pi/3) \rho_\mathrm{m,0} \kcut^{-3}$, the microhalo abundance is heavily suppressed.   

\begin{figure}
 \resizebox{3.4in}{!}
 {
      \includegraphics{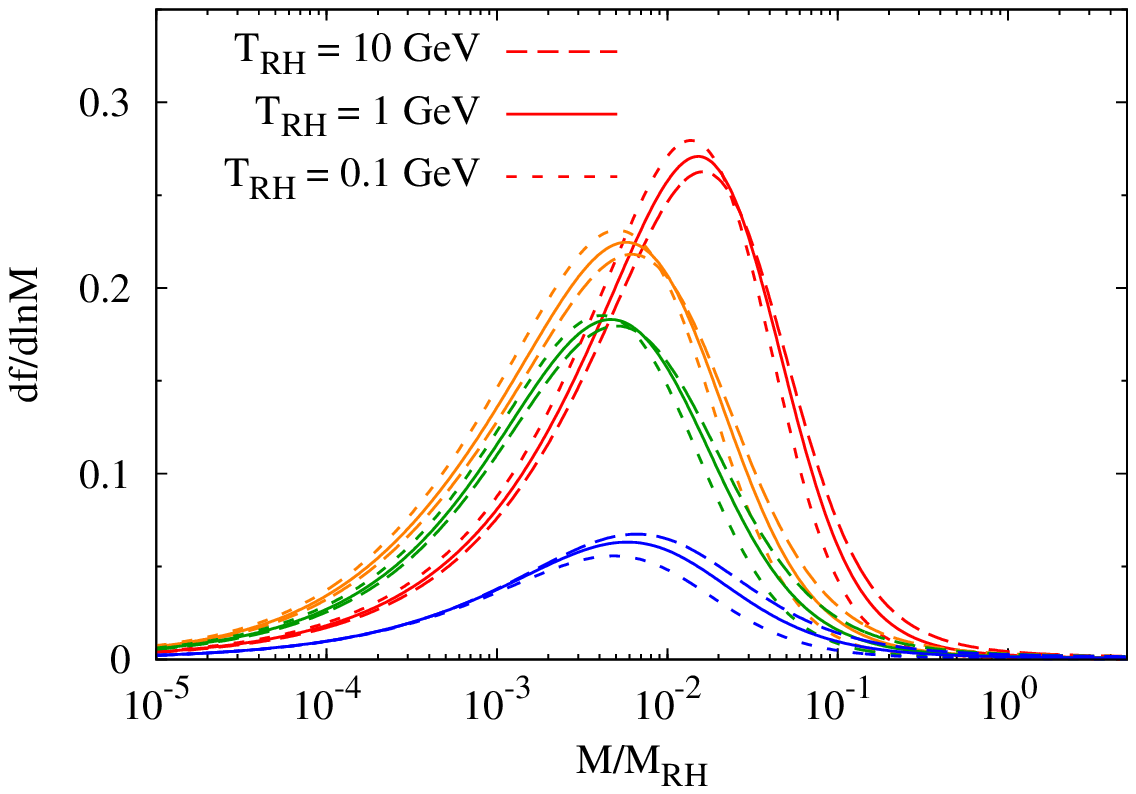}
 }
\caption{The differential bound fraction given by Eq.~(\ref{dfdlnM}) at a redshift $z=50$ for different values of the reheat temperature.  As in Figure \ref{Fig:SigSim}, the four sets of peaks correspond to different values of $\kcut/\krh$: in order of descending maxima, the curves show no cut-off and $\kcut/\krh$ = 40, 20, and 10.}
\label{Fig:BMFSim}
\end{figure}

Nevertheless, Figure \ref{Fig:BMFevol} shows that an EMDE dramatically enhances the microhalo abundance, even if $\kcut$ is as small as $10\krh$.  In the absence of an EMDE, the differential bound fraction inherits the weak mass dependence of the $\sigma(M)$ function.  Therefore, the differential bound fraction for $M>\mdec$ in Figure \ref{Fig:BMFevol}, which only begins to noticeably deviate from zero for $z\lsim 25$, gives an accurate estimate of the differential bound fraction for $M\lsim \mdec$ in the absence of an EMDE.  The onset of standard structure formation at the lower redshifts shown in Figure \ref{Fig:BMFevol} decreases the fraction of dark matter in microhalos with masses less than $\mdec$ as these microhalos are absorbed into larger halos. 

Figure \ref{Fig:BMFevol} shows the bound fraction for $\Trh = 5$ GeV, but the evolution of the bound fraction as a function of $M/\mdec$ is the same for other reheat temperatures, as demonstrated in Figure \ref{Fig:BMFSim} for $z=50$.   This insensitivity to $\Trh$ is not surprising given that the bound fraction is determined by $\sigma(M)$ and, as seen in Figure \ref{Fig:SigSim}, $\sigma(M/\mdec)$ has a very mild dependence on $\Trh$.  Therefore, while the reheat temperature determines the masses of the microhalos generated by an EMDE, it does not significantly change the fraction of the dark matter that is contained in these microhalos, nor does it affect the redshift at which these microhalos form.  The annihilation rates estimated in the next section depend only on the bound fraction and the redshift of microhalo formation, so they are not highly sensitive to the reheat temperature.

\section{Annihilation Signatures}
\label{sec:annihilation}

The annihilation rate for dark matter is
\beq
\Gamma = \frac{\sigv}{2 \mdm^2} \int \rho_\chi^2(\vec{r}) \,\drm^3 \vec{r} \equiv  \frac{\sigv}{2 \mdm^2} J.
\label{annrate}
\eeq
In this expression, $J$ has been defined to characterize the dependence of the annihilation rate on the distribution of the dark matter as opposed to its microphysics.  Since the annihilation rate is proportional to $\rho_\chi^2$, the presence of substructure increases the dark matter annihilation rate, which is often referred to as a substructure boost.  For a halo with mass $M$, the boost factor 
\beq
1+B(M) \equiv \frac{J}{ \int \bar{\rho}_\chi^2(r) \, 4\pi r^2 \drm r}
\label{boostdef}
\eeq
quantifies how the annihilation rate differs from what it would be if the halo had a smooth density profile $\bar{\rho}_\chi(r)$.  In the absence of an EMDE, the boost factor is often calculated by extrapolating the subhalo mass function determined by N-body simulations down to the scale of the smallest microhalos, as determined by the cut-off in the matter power spectrum \cite[e.g.][]{SKB07, KKK10, ZSB10, AD12, Ishiyama14, SP14}.  Most of these analyses predict that $B(M)$ is between 2 and 10 for $M = 10^8 M_\odot$ and between 4 and 40 for $M=10^{12} M_\odot$; Refs.~\cite{SKB07,ZSB10} obtain larger values of $B$ because they extrapolate a power-law relationship between concentration and mass.  The range in possible boost factors stems from uncertainties in both the subhalo mass function and the minimum halo mass.

\begin{figure}
 \resizebox{3.4in}{!}
 {
      \includegraphics{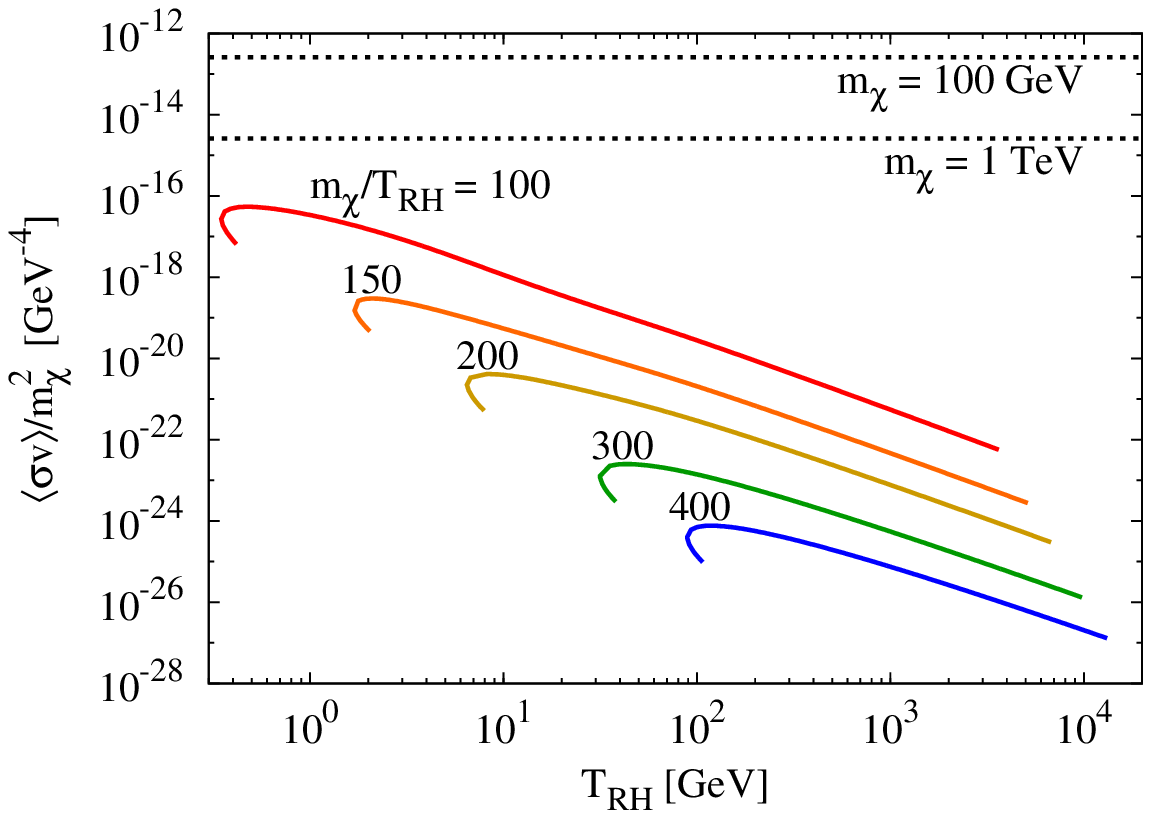}
 }
\caption{The microphysics factor that determines the annihilation rate ($\sigv/\mdm^2$) for freeze-out scenarios that generate the observed dark matter abundance.  For each value of $\mdm/\Trh$ and $\Trh$, $\sigv$ was chosen so that Eq.~(\ref{OmegaFO}) gives $\Omega_\chi h^2 = 0.12$.  As in Figure \ref{Fig:Tfreeze}, only scenarios that obey the unitary condition $\sigv < \mdm^{-2}$ are shown.  The dotted horizontal lines show $\sigv/\mdm^2$ for $\sigv = 3 \times 10^{-26} \sigunits$, which generates the observed dark matter density in the absence of an EMDE.}
\label{Fig:AnnFactor}
\end{figure}

Extrapolating the subhalo mass function is only justifiable if the matter power spectrum is featureless on scales between the resolution limit of the simulations and the cut-off scale.  Since an EMDE significantly enhances the abundance of microhalos with $M_\mathrm{cut}<M<\mdec$, it can generate a boost factor that greatly exceeds the value predicted by extrapolating the subhalo mass function seen in N-body simulations.  However, as seen in Figure \ref{Fig:OmegaDM}, an EMDE also decreases the value of $\sigv$ that generates the observed dark matter density.  Furthermore, Figure \ref{Fig:Tfreeze} indicates that $\mdm \gsim 100 \Trh$ is required to protect the enhanced microhalo abundance generated by an EMDE from the effects of dark matter free-streaming.  Together, these results indicate that the same EMDE scenarios that increase the boost factor will suppress $\sigv/\mdm^2$.   In the absence of an EMDE, $\sigv = 3 \times 10^{-26} \sigunits$ generates the observed dark matter abundance, which implies that 
\beq
\frac{\sigv}{\mdm^2} = \frac{2.6 \times 10^{-13}}{\mathrm{GeV}^4} \left(\frac{100 \, \mathrm{GeV}}{ \mdm} \right)^2.
\eeq
Figure~\ref{Fig:AnnFactor} shows that $\sigv/\mdm^2 < 10^{-16}$ for all EMDE scenarios with $\mdm/\Trh \geq 100$.  Therefore, the overall impact of an EMDE on the dark matter annihilation rate depends on the boost factor from microhalos: is it large enough to compensate for the reduction in $\sigv/\mdm^2$ required to generate the observed dark matter density?

\subsection{The EMDE Boost Factor}
\label{sec:boost}

To estimate the boost factor generated by an EMDE, I will assume that the microhalos that form at high redshift continue to track the dark matter density as larger halos form.  In this case, the number density of microhalos within a halo is determined by the differential bound fraction at high redshift:
\beq
\frac{d n_\mathrm{micro}}{d \ln M_f} = s(r)\frac{\rho_\mathrm{halo}(r)}{M_f} \left. \frac{d f}{d \ln M_f}\right|_{z= z_f},
\label{massfunc}
\eeq
where $M_f$ is the mass of the microhalo at high redshift ($z=z_f$), $\rho_\mathrm{halo}(r)$ is the spherically averaged present-day density of the halo as a function of distance from its center, and $s(r)$ is the fraction of microhalos that survive at that position within the host halo.  This microhalo mass function does not assume that the microhalos retain all of their original mass; due to tidal stripping, the present-day mass of a microhalo is most likely much smaller than $M_f$.   Since $\rho_{\mathrm{m},0}$ was used to relate $M$ and $R$ when computing $\sigma(M)$, both $M_f$ and $\rho_\mathrm{halo}$ should include baryons.  Although the baryons and dark matter are distributed very differently within the halo, this separation will not affect the boost factor because $\rho_\mathrm{halo}$ will be integrated over the volume of the halo in Eq.~(\ref{Jmicro}).  

The redshift $z_f$ should be chosen to be large enough that the integrated bound fraction for $M<\mdec$ has not yet begun to decrease, which indicates that the microhalos are being absorbed into larger halos.  Figure~\ref{Fig:BMFevol} indicates that $z_f \gsim 25$ is required to characterize the EMDE-enhanced microhalo population prior to the onset of standard structure formation.  Beyond this lower limit, the best choice for $z_f$ depends on how the evolution of the bound fraction is interpreted. For example, if $\kcut/\krh = 20$, then Figure \ref{Fig:BMFevol} indicates that microhalos with $M\lsim 0.001 \mdec$ begin to form at $z\simeq 400$, and at $z=50$, most of the dark matter is bound into microhalos with $M \lsim 0.1 \mdec$.  Unfortunately, the Press-Schechter formalism does not tell us about the internal structure of the microhalos: are the microhalos present at $z=50$ smooth halos, in which case $z_f \simeq 50$ would best characterize the microhalo population, or are they bound collections of smaller microhalos that formed at higher redshift?  In this case, a higher value of $z_f$ would give a more accurate prediction of the boost factor.  Given this ambiguity, I will keep $z_f$ as a free parameter until the final evaluation of the boost factor, at which point the implications of its selection will be more transparent.

The halo's $J$ factor, as defined in Eq.~(\ref{annrate}), is the sum of the contribution from the EMDE-generated microhalos and the contribution from the rest of the host halo, which includes all the substructure that would be present in the absence of an EMDE.  If $f_s(r)$ is the present-day fraction of dark matter that is not bound into EMDE-generated microhalos, 
\beq
J = J_\mathrm{micro} + \int_0^R f_s^2(r) \rho_\mathrm{\chi, halo}^2(\vec{r})\, \drm^3 \vec{r},
\label{Jfull}
\eeq
where $\rho_\mathrm{\chi,halo}(\vec{r})$ is the dark matter density of the halo, including all the substructure that is not attributable to the EMDE, and
\begin{align}
J_\mathrm{micro} =& \int_0^{\ln \mdec}\left( \int_0^R J_\mathrm{cl}(M_f) \,\frac{d n_\mathrm{micro}}{d \ln M_f}  \, \drm^3 r \right) d \ln M_f, \\
=& \int_0^{\ln \mdec} \frac{J_\mathrm{cl} (M_f)}{M_f} \,\left.\frac{d f}{d \ln M_f}\right|_{z=z_f}{d \ln M_f} \nonumber \\
&\times\int_0^R s(r) \rho_\mathrm{halo}(r) \,\drm^3 r,  \label{Jmicro}
\end{align}
where $J_\mathrm{cl}(M_f)$ is the $J$ factor for a single microhalo.  In the second line, I have assumed that $J_\mathrm{cl}$ does not depend on the microhalo's position within the host halo.  Although the microhalo's position will affect how much mass it loses due to tidal stripping, $J_\mathrm{cl}$ is not greatly affected by these spatial variations because it is dominated by the contribution from the microhalo's innermost region.

Numerical simulations of microhalo formation in the absence of an EMDE \cite{DMS05, IME10, AD12, Ishiyama14} find that microhalos with masses just above the free-streaming scale have extremely cuspy density profiles: $\rho(r)\propto r^{-\gamma}$ with $1.3<\gamma<1.5$ in the innermost regions.  However, larger microhalos and microhalos that form in the absence of a cut-off in the matter power spectrum have shallower cusps with $\gamma \simeq 1$ in the inner region; the density profiles of these microhalos are well fit by a Navarro-Frenk-White (NFW) profile.   Mergers between microhalos of similar mass are thought to lead to the softer NFW profile \cite{Ishiyama14}.  The evolution of the bound fraction shown in Figure \ref{Fig:BMFevol} indicates that microhalo formation after an EMDE is strongly hierarchical, so it is reasonable to expect that microhalos experience several major mergers shortly after their formation.  To be conservative, I adopt an NFW density profile for the dark matter within the microhalos:
\beq
\rho_\chi(r)= \frac{\rho_s}{(r/r_s)(1+r/r_s)^2}.
\eeq
The scale radius $r_s$ is parametrized by the halo's concentration: $c = r_\mathrm{200}/r_s$, where $r_\mathrm{200}$ is the radius within which the average matter density is 200 times the critical density ($\rho_{200} \equiv 200 \rho_\mathrm{crit}$).  The normalization of the NFW profile is determined by $\rho_{200}$ and the concentration:
\beq
\rho_s = f_\chi \frac{\rho_{200}}3 \frac{c^3}{\ln (1+c) -c/(1+c)},
\eeq
where $f_\chi \equiv \rho_\chi/\rho_m$ is the dark matter fraction.  If $M_{200}$ is the total mass enclosed within $r_{200}$, this value of $\rho_s$ sets the dark matter mass enclosed within $r_{200}$ to be $f_\chi M_{200}$.

Since the critical density decreases with time, $r_\mathrm{200}$ is roughly proportional to $(1+z)^{-1}$ while the Universe is matter dominated.  Since $r_s$ is expected to be constant, $c_{200} \propto (1+z)^{-1}$ \cite{BKS01,WBP02}.  Simulated microhalos have concentrations between 2 and 3 at a redshift of $\sim$30 \cite{AD12, Ishiyama14}.  It is unknown how an EMDE would affect microhalo concentrations, but since the earliest forming microhalos in these simulations have the highest concentrations \cite{Ishiyama14}, the microhalos generated by an EMDE will probably be even more concentrated.  Without an EMDE, the first microhalos form at a redshift of 60; if concentration scales with formation redshift \cite{BKS01, WBP02, MDB08} and is otherwise unaffected by the EMDE, then the concentration of a microhalo that forms at a redshift $z_f$ will be
\beq
c_{200}(z) \sim 3\left(\frac{1+z_f}{60}\right)\left(\frac{30}{1+z}\right).
\label{con}
\eeq

Although the NFW density profile diverges at the center of the halo, the central density is limited by both the annihilation rate and the phase-space density, which cannot increase after the dark matter particles decouple \cite{GT79}.   If the annihilating dark matter at the center of the microhalo is not replenished by infalling particles, the maximum density that can be achieved is $\rho_\mathrm{max} = \mdm/(\sigv t_0)$, where $t_0$ is the age of the microhalo \cite[e.g.][]{BDE14}.  The radius at which the NFW profile exceeds this density is the annihilation core radius $r_c$.  If $r_c \ll r_s$ then
\beq
\frac{r_c}{r_s} \simeq \frac{\sigv}{\mdm} t_0 \rho_s.
\eeq
For dark matter particles that freeze out prior to reheating and generate the observed dark matter abundance, $\sigv/\mdm$ increases as $\mdm/\Trh$ decreases.  If \mbox{$\mdm/\Trh > 50$}, then \mbox{$\sigv/\mdm < 8 \times 10^{-13}$ GeV$^{-3}$}.  Therefore, a microhalo that formed at very high redshift will have
\beq
\frac{r_c}{r_s} \lsim 10^{-7} h^2 \frac{(1+z)^3c^3}{10^8},
\eeq
where $(1+z)^3c^3 \lsim 10^8$ follows from Eq.~(\ref{con}) for microhalos that form after $z = 400$.  For microhalos with NFW profiles, the presence of a constant-density core decreases $J_\mathrm{cl}$ by less than 10\% if $r_c/r_s < 0.05$, so the core generated by annihilating dark matter will not significantly affect the annihilation rate within a microhalo.

For the small values of $\sigv/\mdm$ required to thermally generate dark matter during an EMDE, the upper limit on the phase-space density set by the phase-space density when the dark matter kinetically decoupled imposes a stronger constraint on the central density of the microhalo \cite{GT79, DH01,IME10,BDE14}.  If dark matter  kinetically decouples during an EMDE, the phase-space density $Q \simeq \rho/\langle v^2 \rangle^{3/2}$ at decoupling is 
\begin{align}
Q_\mathrm{kd} &\simeq \rho_{\chi,0} a_\mathrm{kd}^{-3} \left(\frac{\mdm}{\Tkd}\right)^{3/2}, \\
&\gsim  \rho_{\chi,0}  \left(\frac{\mdm}{\Trh}\right)^{3/2} \left[\frac{\gstarRH \Trh^3}{3.91 T_0^3}\right]\left(\sqrt{\frac{5}{2}}\frac{\Tkds^2}{\Trh^2}\right)^{13/2}\!\!\!\!,\\
&\gsim \frac{10^{19} M_\odot}{\mathrm{pc^3 \, (km/s)^3}}\left[\frac{\Trh}{80 \, \mathrm{MeV}}\right]^3 \left[\frac{\mdm/\Trh}{50}\right]^{3/2}\left[\frac{\Tkds}{2 \Trh}\right]^{13}\!\!\!\!,
\end{align}
where $T_0 = 2.7 $K is the present-day radiation temperature and Eq.~(\ref{Tkd}) has been used to put a lower bound on $\Tkd/\Trh$.  By Liouville's theorem, the collisionless evolution of the dark matter particles after decoupling cannot increase the phase-space density, so the phase-space density inside the microhalo cannot exceed $Q_\mathrm{kd}$.  The circular velocity $v_c(r) = \sqrt{GM(r)/r}$ can be used to estimate $\langle v^2\rangle$ within the halo; if $f_\chi\rho \simeq \rho_s (r_s/r)$, as is the case for an NFW profile at radii much smaller than $r_s$,
\beq
Q \simeq \frac{f_\chi^{3/2}}{(2\pi G)^{3/2}\sqrt{\rho_s r_s^6}} \left(\frac{r}{r_s}\right)^{-2.5}.
\eeq
 Setting $Q=Q_\mathrm{kd}$ defines the radius of the constant-density core required by Liouville's theorem:
\beq
\frac{r_c}{r_s} \lsim 0.04 \left[\frac{M/\mdec}{10^{-4}}\right]^{-2/5}\left[\frac{1+z_f}{400}\right]^{3/5}\left[\frac{\Tkds}{2\Trh}\right]^{-26/5},
\eeq
for $\mdm/\Trh>50$.  Figure~\ref{Fig:BMFevol} shows that most microhalos that form after $z=400$ have masses greater than $0.00005\mdec$.  Furthermore, $\Tkds/\Trh > 2$ is required to make $\kcut/\krh >10$, so $r_c \lsim 0.05 r_s$ for all cases of interest.  Therefore, the constant-density core required by Liouville's theorem does not reduce $J_\mathrm{cl}$ by more than 10\%, provided that we restrict ourselves to $z_f < 400$.   The core's impact on $J_\mathrm{cl}$ decreases even further as $\mdm/\Trh$ is increased beyond 50.

The annihilation rate within a microhalo is also rather insensitive to the removal of mass from its outer regions; truncating the microhalos' NFW density profile at $r_s$ reduces $J_\mathrm{cl}$ by only 10\% if $c<3$.  Simulations of subhalos with NFW profiles indicate that the subhalo is completely destroyed if its tidal radius is less than twice its scale radius $r_s$ \cite{HNT03}.  Therefore, tidal stripping will most likely destroy a microhalo before it significantly reduces $J_\mathrm{cl}$.   If the gravitational potentials of the host halo and the microhalo are approximated as being generated by point masses, the microhalo's tidal radius is determined by the Roche limit: if $R$ is the microhalo's distance from the center of the host,
\beq
r_t = \left[\frac{M_\mathrm{micro}(r_t)}{2M_\mathrm{host}(R)}\right]^{1/3} R,
\label{rtdef}
\eeq
which implies that the average density of the microhalo within $r_t$ is twice the average density of the host within $R$.  Accounting for the extended mass profile of the host increases the tidal radius \cite{HNT03}, so the Roche limit provides a simple and conservative estimate of $r_t$.  To compute $M_\mathrm{host}(R)$, I assume that the host halo has an NFW profile with the concentration-mass relation provided in Ref. \cite{SP14}, which is based on an analysis of halo concentrations in N-body simulations \cite{PKC12}.  With this host profile, a microhalo with $c \geq 2$ and $z_f \geq 50$ has $r_t > 2r_s$ if $R>1\, \mathrm{kpc}$ and  $M_{200}<10^{14} M_\odot$.  As the mass of the host halo decreases, $r_t/r_s$ at a given position decreases, so microhalos can survive deeper within smaller hosts.  For all host masses, $r_t/r_s > 2$ if $R> \sqrt{R_s/(100\, \mathrm{kpc})}\,\mathrm{kpc}$, where $R_s$ is the scale radius of the host.  If the host mass is greater than $10^7M_\odot$, less than 1\% of the host halo's mass lies within a radius of $\sqrt{R_s/(100\, \mathrm{kpc})}\,\mathrm{kpc}$, so the tidal destruction of the microhalos in the innermost region of the host does not significantly affect their overall abundance. 

The Earth-mass microhalos that form in cosmological scenarios without an EMDE are more vulnerable to stripping by stellar encounters than by the host halo \cite{DMS05, ZHATS07, BDE06, GG07, GGMDS07, AZ07, SKM10, IME10}.  Although these microhalos are expected to lose a considerable fraction of their mass due to interactions with stars, some high-density cores should survive \cite{BDE06, SKM10, IME10}.  For example, simulations of microhalo-star scatterings indicate that the star must pass within 0.02 pc of the microhalo's center to significantly reduce its density within its scale radius and that microhalos should be able to survive outside the host's inner kpc \cite{IME10}.  Furthermore, the microhalos that are generated by an EMDE form far earlier than these standard microhalos, so they have much higher central densities and are less likely to be destroyed by stars.  The microhalo survival rate in the Solar neighborhood increases sharply with the microhalo's mean density \cite{Berezinsky10}: whereas a microhalo that formed at $z=50$ has only a 10\% chance of surviving, the survival rate increases to 85\% for a microhalo that formed at $z=200$.  Although the impact of stellar encounters on EMDE-generated microhalos certainly warrants further investigation, it seems reasonable to assume that most of these microhalos survive outside the innermost kiloparsec of a galactic halo.  

If we conservatively truncate the microhalo density profile at $r=r_s$ and neglect the slight reduction in $J_\mathrm{cl}$ due to a constant-density core,
\begin{align}
J_\mathrm{cl}(M,z) &= \frac76\pi \rho_s^2 r_s^3, \\
&= f_\chi^2\frac{M}3 \rho_{200}(z)c(z,z_f)^3 \frac{7/24}{[\ln (1+c)-c/(1+c)]^2},  \label{Jcl}
\end{align}
where $z_f$ is the formation redshift for a microhalo with mass $M = M_{200}(z)$. (Although $M_{200}$ is not precisely the virial mass predicted by top-hat collapse, the difference is not significant, and using $\rho_{200}$ as the virial density eases comparisons with other microhalo investigations.)  As a dark matter halo accretes mass, the product $\rho_s r_s$ remains roughly constant, while $r_s$ grows approximately logarithmically with concentration \cite{WBP02, RHH07}.  If the EMDE microhalos evolve in the same way prior to their absorption into larger halos, then $J_\mathrm{cl}$ for a given halo will increase with time due to the increasing truncation radius.  After the microhalos become subhalos, simulations of larger satellites indicate that $r_s$ remains roughly constant while $\rho_s$ decreases as the subhalo loses mass \cite{HNT03}.  However, since the microhalos are most likely accreted relatively soon after their formation (while their concentrations are still low) most of their mass will be within 2$r_s$ and may not be stripped.  Therefore, the possible reduction in $\rho_s$ due to stripping will be neglected, as will the possible increase in $r_s$ prior to the microhalo's absorption.  $J_\mathrm{cl}(M)$ will be evaluated at the same high redshift as the microhalo mass function.

At a given redshift, the Press-Schechter mass function includes halos that formed at a range of higher redshifts, with smaller halos forming earlier on average.  Therefore, $z_f$ and by extension $c$ should be functions of $M$.  Since $c \propto (1+z_f)$, increasing a halo's formation redshift will increase $J_\mathrm{cl}(M)$.  For example, if $J_\mathrm{cl}$ is evaluated at a redshift of 50 and Eq.~(\ref{con}) is used to evaluate the concentration, then $J_\mathrm{cl}$ increases by a factor of 18 as $z_f$ increases from 50 to 400. Therefore, ignoring this variation between microhalos and treating all microhalos as if they formed shortly before the bound fraction is evaluated provides a lower bound on the boost factor.  
With this approximation, the concentration is independent of microhalo mass, and inserting Eq.~(\ref{Jcl}) into Eq.~(\ref{Jmicro}) yields
\begin{align}
J_\mathrm{micro} \gsim & \frac{(7/72) \rho_{200}(z_f)c^3f_\chi^2}{[\ln (1+c)-c/(1+c)]^2} \int_0^{\ln \mdec}\!\!\! \,\left.\frac{d f}{d \ln M}\right|_{z_f}{d \ln M} \nonumber \\
&\times\int_0^R s(r) \rho_\mathrm{halo}(r) \,\drm^3 r.
\end{align}
I will assume that all microhalos have $c=2$ at a redshift $z_f$ because this is the lowest concentration seen in microhalo simulations \cite{AD12, Ishiyama14}.

\begin{figure}
 \resizebox{3.4in}{!}
 {
      \includegraphics{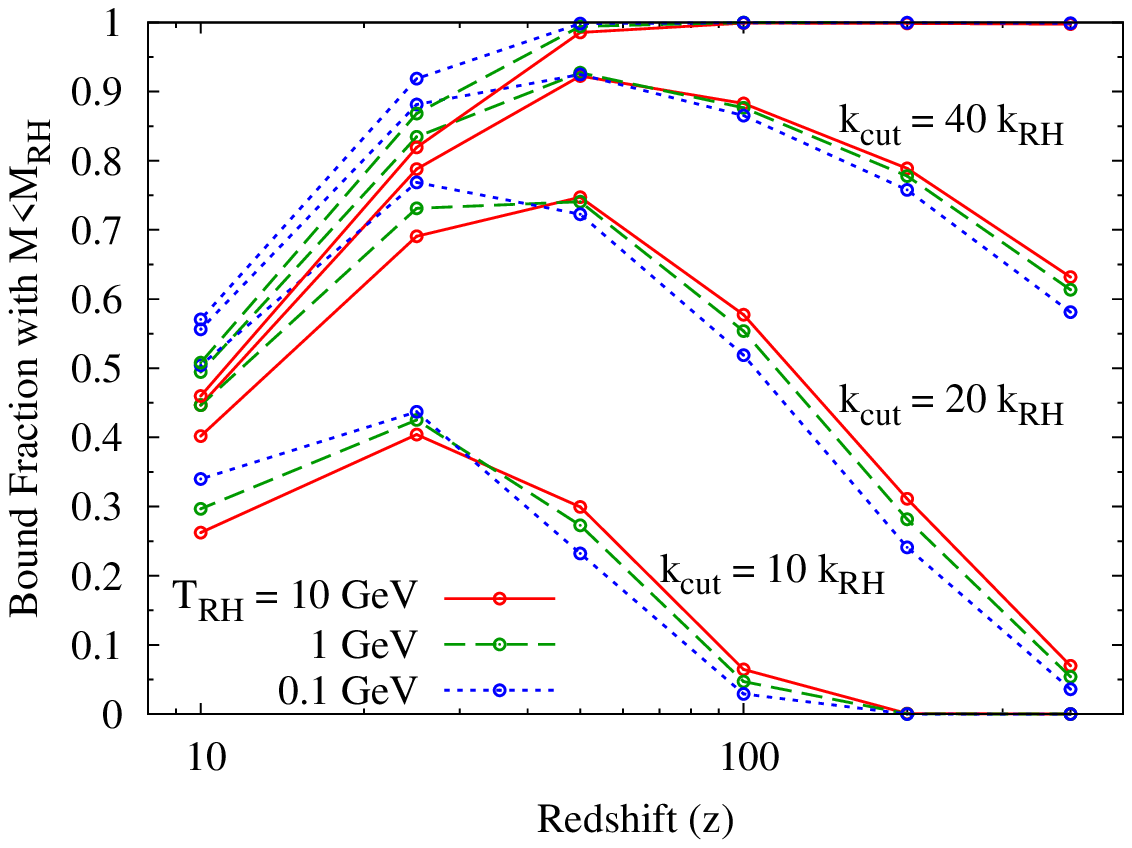}
 }
\caption{The fraction of dark matter bound into halos with masses less than $\mdec$ as a function of redshift.  The top set of curves does not have a small-scale cut-off in the matter power spectrum; in this case, all of the dark matter is bound into microhalos at arbitrarily high redshifts.  The lower three sets of curves have power spectra that are exponentially suppressed for $k > \kcut$.  In all cases, the bound fraction decreases at low redshifts when some microhalos are absorbed into halos with masses greater than $\mdec$.  For comparison, in the absence of an EMDE, the \mbox{$10^{-20}M_\odot<M<\mdec$} bound fraction at $z\geq 50$ is less than 0.05 for all possible reheat temperatures.}
\label{Fig:BFtot}
\end{figure}

If the microhalo survival rate $s(r)$ is taken to be 0 inside some radius $R_\mathrm{min}$ and 1 outside this radius, then 
\beq
J_\mathrm{micro} \gsim 4f_\chi^2\rho_{200}(z_f) f_\mathrm{tot}(z_f)M_\mathrm{halo}(r>R_\mathrm{min}),
\label{JmicroSim}
\eeq
where $f_\mathrm{tot}(z_f)$ is the fraction of dark matter bound into microhalos with $M < \mdec$ at a redshift $z_f$.  Figure \ref{Fig:BMFSim} indicates that $f_\mathrm{tot}$ will not depend strongly on $\Trh$ for a fixed value of $\kcut/\krh$.  Figure \ref{Fig:BFtot} shows $f_\mathrm{tot}(z_f)$ for reheat temperatures of 0.1 GeV, 1 GeV, and 10 GeV and confirms this insensitivity to $\Trh$.   Figure \ref{Fig:BFtot} also shows how $f_\mathrm{tot}(z_f)$ depends on $\kcut/\krh$: as expected from Figure \ref{Fig:BMFevol}, decreasing $\kcut/\krh$ delays the formation of microhalos and decreases their abundance.   Therefore, while $J_\mathrm{micro}$ has a very mild direct dependence on the reheat temperature, it is very sensitive to changes in the ratio of the decoupling temperature to the reheating temperature.  Since $\rho_{200}(z_f) \propto (1+z_f)^3$, increasing the redshift at which $f_\mathrm{tot}(z_f)$ is nonzero will dramatically increase $J_\mathrm{micro}$.  

Figure \ref{Fig:BFtot} also provides guidance regarding the optimal value of $z_f$.  The lower bound on $J_\mathrm{micro}$ given in Eq.~(\ref{JmicroSim}) was derived assuming that all the microhalos present at $z_f$ formed near that redshift.  Since earlier-forming microhalos have higher densities, Eq.~(\ref{JmicroSim}) depends on the product $\rho_{200}(z_f) f_\mathrm{tot}(z_f)$, which is not maximized at the same redshift that maximizes the bound fraction.  For example, if $\kcut = 20 \krh$, then only about 5\% of the dark matter is bound into microhalos at a redshift of 400, whereas about 75\% of the dark matter is bound into microhalos at a redshift of 50, and yet, $J_\mathrm{micro}(z_f = 400) \simeq 30 J_\mathrm{micro}(z_f=50)$.  Therefore, if the microhalos that are present at $z=400$ survive today, their contribution to $J$ is greater than the contribution from the later-forming microhalos.  Unfortunately, the fate of the earliest forming microhalos is unknown; N-body simulations of microhalo formation after an EMDE are required to determine whether they survive their absorption into the larger microhalos that form at slightly lower redshifts.  If they are destroyed, then $z_f$ should be chosen to maximize $f_\mathrm{tot}(z_f)$, as this redshift gives the most accurate characterization of the microhalo population generated by an EMDE.  Conversely, if they survive as subhalos within microhalos, then $z_f$ should be chosen to maximize Eq.~(\ref{JmicroSim}).  In either case, Eq.~(\ref{JmicroSim}) provides a lower bound on $J$; choosing a large value for $z_f$ ignores the contribution of later-forming microhalos, while choosing a small value for $z_f$ ignores the fact that some of the microhalos at that redshift formed much earlier and have higher internal densities.

\begin{figure*}
 \centering
\begin{minipage}{0.5\textwidth}
\centering
 \resizebox{3.4in}{!}
 {
      \includegraphics{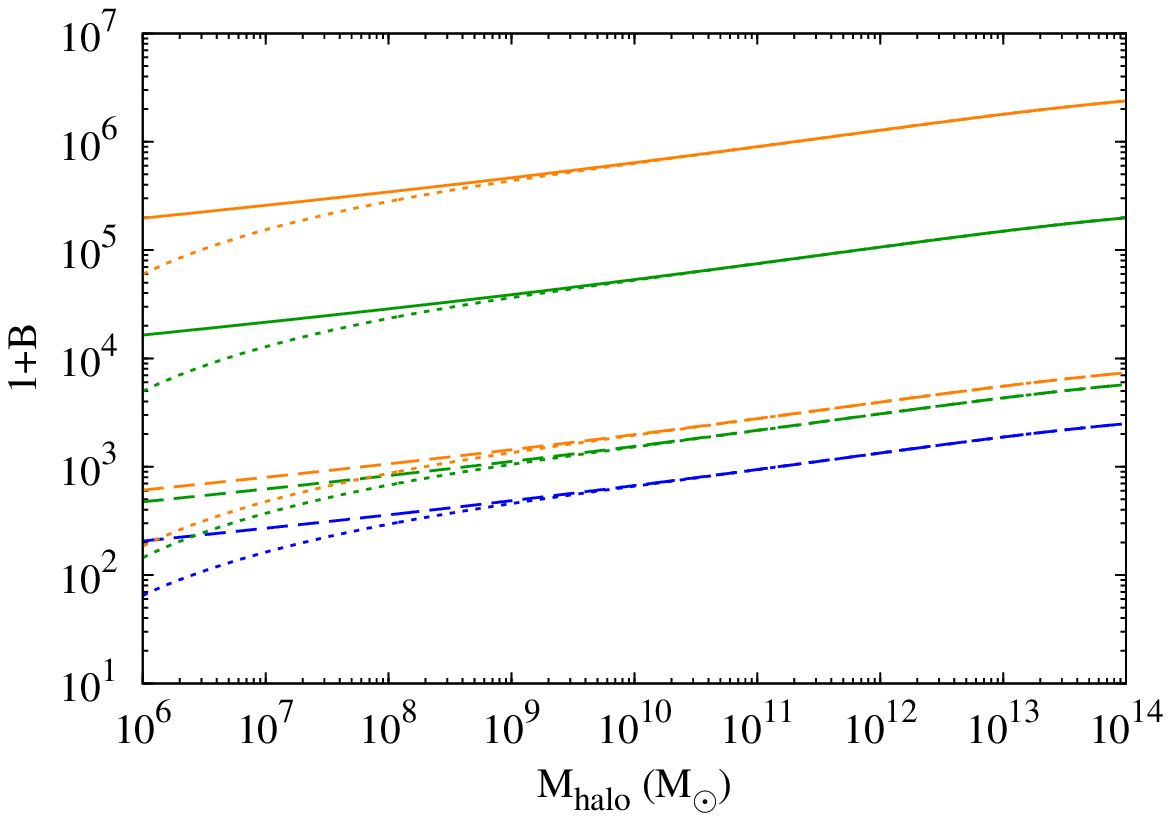}
 }
\end{minipage}%
\begin{minipage}{0.5\textwidth}
\centering
 \resizebox{3.4in}{!}
{
      \includegraphics{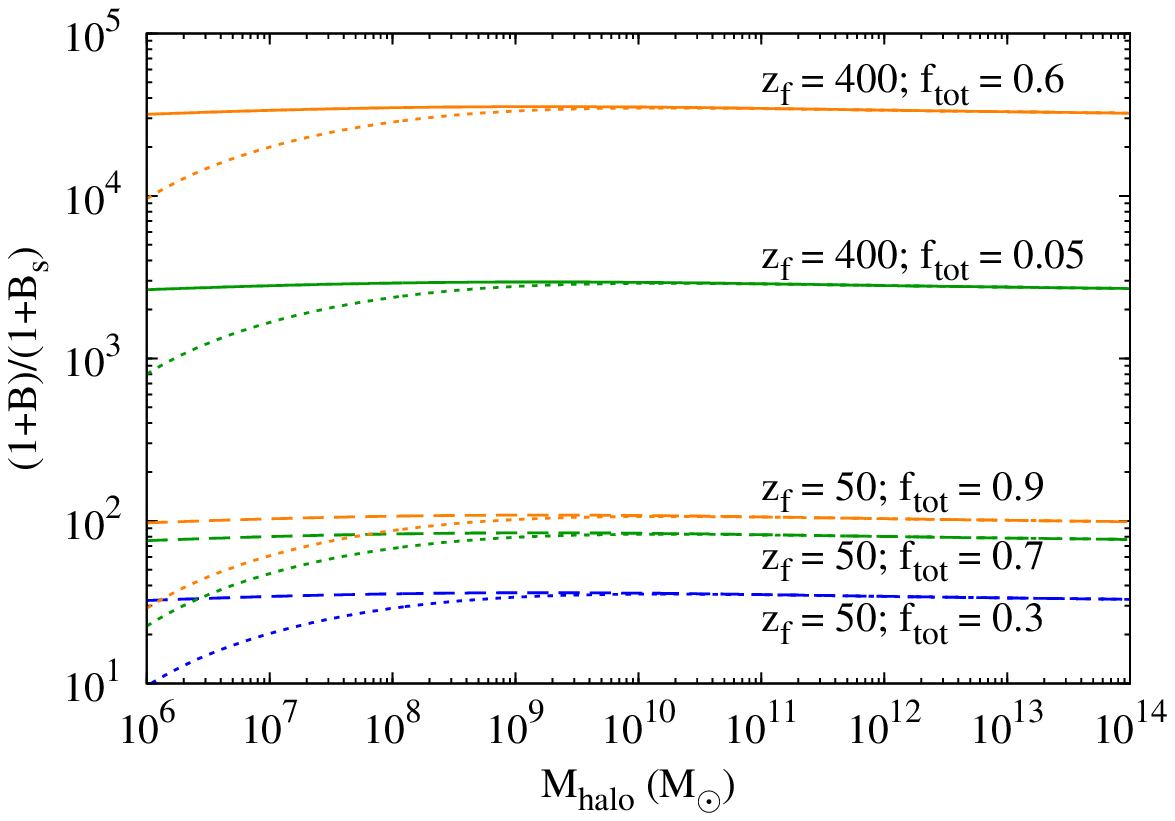}
 }
\end{minipage}%
\caption{The boost factor and relative boost factor generated by an EMDE.  The left panel shows the boost factor as defined by Eq.~(\ref{boostdef}), while the right panel shows this boost factor divided by the boost factor from standard subhalos with masses greater than $10^{-12} M_\odot$ \cite{SP14}.  Both panels show the same five combinations of $z_f$ and $f_\mathrm{tot}$ as labeled in the right panel. For each value of $z_f$, the largest value of $f_\mathrm{tot}$ shown corresponds to $\kcut/\krh = 40$, while the second-largest values of $f_\mathrm{tot}$ corresponds to $\kcut/\krh = 20$.  Finally, the smallest boost corresponds to $\kcut/\krh = 10$; microhalos are not present at a redshift of 400 in this scenario.
For each value of $z_f$ and $f_\mathrm{tot}$, the dotted curve assumes that the EMDE-generated microhalos are destroyed within the innermost kiloparsec of the host halo, while the solid ($z_f = 400$) and dashed ($z_f = 50$) curves take $R_\mathrm{min} = \sqrt{R_s/(100 \,\mathrm{kpc}) }$ kpc.}
\label{Fig:Boost}
\end{figure*}

The bound fraction $f_\mathrm{tot}(z_f)$ is not directly related to the fraction $f_s(r)$ of dark matter that is not contained in microhalos today, which appears in the expression for $J$ given by Eq.~(\ref{Jfull}).  First, if all microhalos are destroyed within $R_\mathrm{min}$ of the center of the halo, then $f_s(r<R_\mathrm{min})$ will equal one.  Second, the microhalos will lose mass as they are tidally stripped within the larger halos.  A halo with a concentration of 1.5 that is stripped to $r_s$ retains only 61\% of its original mass, and the mass loss increases as the concentration increases.  Therefore, \mbox{$f_s(r) \gsim 1-0.6 f_\mathrm{tot}(z_f)$}.   If $f_s$ is assumed to be constant for $r>R_\mathrm{min}$ and 1 for $r<R_\mathrm{min}$, 
\beq
J = J_\mathrm{micro} + J_\mathrm{halo}(r\!<\!R_\mathrm{min}) + f_s^2 \left[J_\mathrm{halo} - J_\mathrm{halo}(r\!<\!R_\mathrm{min})\right],
\label{Jtot}
\eeq 
where $J_\mathrm{halo}$ includes the boost factor $B_s(M_h)$ from subhalos with $M>\mdec$:
\begin{align}
J_\mathrm{halo} &= [1+B_s(M_h)] 4\pi \int_0^{\infty} r^2 \bar{\rho}^2_\chi(r) \drm r,\\
&= [1+B_s(M_h)] \frac{f_\chi^2 M_h\rho_{200}c_h^3}{9[\ln (1+c_h)-c_h/(1+c_h)]^2}.
\end{align}
The last line assumes that the halo has an NFW profile with $M_{200}=M_h$, scale radius $R_s$, and concentration $c_h$.  The concentration-mass relation provided in Ref. \cite{SP14} successfully matches the concentrations of simulated halos with masses ranging from $10^{-5} M_\odot$ to $10^{15} M_\odot$, and I will continue to use it to evaluate the present-day concentration of the host halo.  Since subhalos with $M>\mdec$ are less dense than the EMDE-generated microhalos, they should not survive at radii smaller than $R_\mathrm{min}$, which implies that 
\begin{align}
J_\mathrm{halo}(r\!<\!R_\mathrm{min}) &= 4\pi \int_0^{R_\mathrm{min}} r^2 \bar{\rho}^2_\chi(r) \drm r,\\
&= \frac{f_\chi^2M_h\rho_{200}c_h^3}{3[\ln (1+c_h)-c_h/(1+c_h)]^2}{\cal J}\left(\frac{R_\mathrm{min}}{R_s}\right),
\end{align} 
where 
\beq
{\cal J}(y) \equiv \frac{y^3 + 3y^2+3y}{3(y+1)^3}.
\eeq

The total boost factor generated by an EMDE is obtained by dividing Eq.~(\ref{Jtot}) by $4\pi \int_0^{\infty} r^2 \bar{\rho}^2_\chi(r) \drm r$:
\begin{widetext}
\beq
1+B \gsim 36 f_\mathrm{tot}(z_f)\left[\frac{\rho_{200}(z_f)}{\rho_{200}}\right]\left[\frac{M_h(r>R_\mathrm{min})}{M_h}\right] \frac{[\ln (1+c_h)-c_h/(1+c_h)]^2}{c_h^3} + [1+B_s(M_h)]f_s^2+3{\cal J}\left(\frac{R_\mathrm{min}}{R_s}\right){(1-f_s^2)}.
\label{boostfinal}
\eeq
\end{widetext}
Figure \ref{Fig:Boost} shows $1+B$ as a function of $M_h$, along with the relative boost from an EMDE: $(1+B)/(1+B_s)$.  In both cases, $B_s(M_h)$ was evaluated using the fitting function in Ref.~\cite{SP14} for the boost factor from subhalos with masses larger than $10^{-6} M_\odot$ and a mass function $dN/dm \propto m^{-2}$ multiplied by a factor of 2.5 to match their results for the boost factor with a minimum subhalo mass of $10^{-12} M_\odot$.   As seen in Figure \ref{Fig:AnnFactor}, the most promising scenarios for detection have reheat temperatures between 350 MeV and 10 GeV, which implies that $\mdec$ is between $10^{-9} M_\odot$ and $10^{-14} M_\odot$.  Since $B(M)$ changes by only a factor of 2.5 when the minimum subhalo mass increases by six orders of magnitude, using $M_\mathrm{min} = 10^{-12} M_\odot$ instead of $M_\mathrm{min} = \mdec$ will not significantly affect the boost factor for these reheat temperatures.  In Figure \ref{Fig:Boost}, $f_s$ has been set to $1-0.6 f_\mathrm{tot}(z_f)$, but since the Eq.~(\ref{boostfinal}) is dominated by the first term for the cases shown, this choice has very little impact. 

In Figure \ref{Fig:Boost}, $z_f$ and $f_\mathrm{tot}$ are chosen to correspond to optimistic and pessimistic scenarios for $\kcut/\krh = 40$ and $\kcut/\krh = 20$.  (If there is no small-scale cut-off in the matter power spectrum, an EMDE will generate microhalos at arbitrarily high redshifts, which will lead to arbitrarily large boost factors.)  In the optimistic case, $z_f=400$ is chosen to maximize Eq.~(\ref{Jmicro}), while in the pessimistic case, the microhalos present at high redshift are assumed to be destroyed and $z_f$ is taken to be 50.  Figure \ref{Fig:Boost} also shows $z_f = 50$ for $\kcut/\krh = 10$; in this case, choosing a higher $z_f$ does not significantly increase the boost factor.   Figure \ref{Fig:Boost} shows that larger halos have larger absolute boost factors, but the relative boost from an EMDE is nearly independent of mass if $R_\mathrm{min}$ is small enough that $M_h(r>R_\mathrm{min}) \simeq M_h$.  The origin of this behavior can be seen in Eq.~(\ref{boostfinal}): larger halos have lower concentrations than smaller halos, which means that high-density microhalos have a larger impact on the dark matter annihilation rate.  However, when standard subhalos are included, they dominate $J_\mathrm{halo}$.  Consequently, the relative boost from an EMDE is attributable to the microhalos' enhancement of the annihilation rate within the subhalos, which does not depend on the mass of the host halo.  

\subsection{Detection outlook}
\label{sec:detect}

For dark matter particles with masses less than a TeV, the most stringent constraints on the dark matter annihilation rate come from observations of dwarf spheroidal galaxies (dSphs) by the \emph{Fermi} Large Area Telescope (Fermi-LAT).  The Fermi-LAT Collaboration has recently updated these constraints using six years of observations of fifteen dSphs \cite{FermidSphs15}.  For dark matter annihilating into $b$ quarks, $\sigv > 2 \times 10^{-26} \sigunits$ is excluded at 95\% confidence level (CL) for $\mdm = 100$ GeV, and $\sigv > 2 \times 10^{-25} \sigunits$ is excluded at 95\% CL for $\mdm = 1$ TeV.   Slightly weaker constraints are found if dark matter annihilates into $\tau$ leptons.
For particles heavier than 1 TeV,  air Cherenkov telescopes provide more stringent constraints on $\sigv$.  For $\mdm = 10$ TeV,  MAGIC observations of Segue 1 \cite{Magic14} and H.E.S.S. observations of five dSphs \cite{HESSdSphs} exclude $\sigv > 10^{-23} \sigunits$ at 95\% CL if dark matter annihilates into $\tau$ leptons, and observations of the Galactic center by H.E.S.S. exclude $\sigv > 10^{-24} \sigunits$ at 95\% CL if dark matter annihilates into $b$ quarks \cite{HESSGC11}.  However, these H.E.S.S. observations are confined to a region within 150 pc of the Galactic center, which is expected to be devoid of microhalos.  Based solely on observations of dSphs, Fermi-LAT provides the strongest constraints on heavy dark matter particles annihilating into $b$ quarks, excluding $\sigv > 4 \times 10^{-24} \sigunits$ at 95\% CL for $\mdm = 10$ TeV \cite{FermidSphs15}.

None of these constraints on $\sigv$ from dSphs include a boost factor from substructure; given the uncertainty in the boost factor, it is considered more conservative to omit it entirely.  Unfortunately, the boost factors shown in Figure \ref{Fig:Boost} cannot be readily applied to observations of dSphs because these observations are confined to the dSphs' central regions.  The Fermi-LAT analysis only considered emission from within $0.5^\circ$ of the dSphs' centers, which corresponds to a radius of 200 pc for Segue 1 and 280 pc for Ursa Major, the two nearest dSphs.  If these galaxies have NFW profiles with the concentration-mass relation given by Ref. \cite{SP14}, then microhalos are expected to survive at radii greater than 30 or 40 parsecs [$R_\mathrm{min} = \sqrt{R_s/(100 \,\mathrm{kpc})}$ kpc] in these systems, so an EMDE will enhance the annihilation rate within this region. However, $J_\mathrm{micro}$ is proportional to the halo mass, and only 15\% of the mass of Ursa Major and 25\% of the mass of Segue 1 lies within $0.5^\circ$ of their centers.  Meanwhile, restricting $J_\mathrm{halo}$ to this region reduces $J_\mathrm{halo}$ by less than 7\%.  Therefore, Fermi-LAT's assumption of a limited annihilation region will reduce the microhalo boost factor by the fraction of mass that is included, and the boost factors shown in Figure \ref{Fig:Boost} should be reduced by roughly a factor of 10 when applied to dSphs.

This reduction in the boost factor for dSph observations makes it worthwhile to consider the nominally weaker constraints from Fermi-LAT's observations of the isotropic gamma-ray background (IGRB) \cite{FermiIGRB15}.  Conservative constraints are derived by assuming that the IGRB originates solely from dark matter annihilations.   The 95\% CL upper limits on $\sigv$ for annihilation into $b$ quarks are $7\times10^{-25}\sigunits$, $3\times10^{-24} \sigunits$, and $5\times10^{-24} \sigunits$ for $\mdm = 100$ GeV, 1 TeV and 10 TeV, respectively.  The upper limits for annihilation to $\tau$ leptons differ by less than an order of magnitude.  To account for substructure, Ref. \cite{FermiIGRB15} uses the same $B_s(M)$ function from Ref.~\cite{SP14} as I do, but with $M_\mathrm{min} = 10^{-6} h^{-1} M_\odot$.  Since the annihilation signal is proportional to the boost factor, decreasing $M_\mathrm{min}$ to $10^{-12} M_\odot$ would decrease all these upper limits on $\sigv$ by a factor of $\sim$2/5.  Ref. \cite{FermiIGRB15} also provides an estimate of Fermi-LAT's sensitivity reach by constraining a possible dark matter annihilation signal that lies on top of a simple model for the contribution to the IGRB from other astrophysical sources.  These sensitivity bounds forecast the constraints on $\sigv$ that could be derived from the IGRB if the astrophysical background were fully understood. For $\mdm\lsim1$ TeV, these upper limits are about an order of magnitude stronger than the conservative upper bounds on $\sigv$.

\begin{figure}
 \resizebox{3.4in}{!}
 {
      \includegraphics{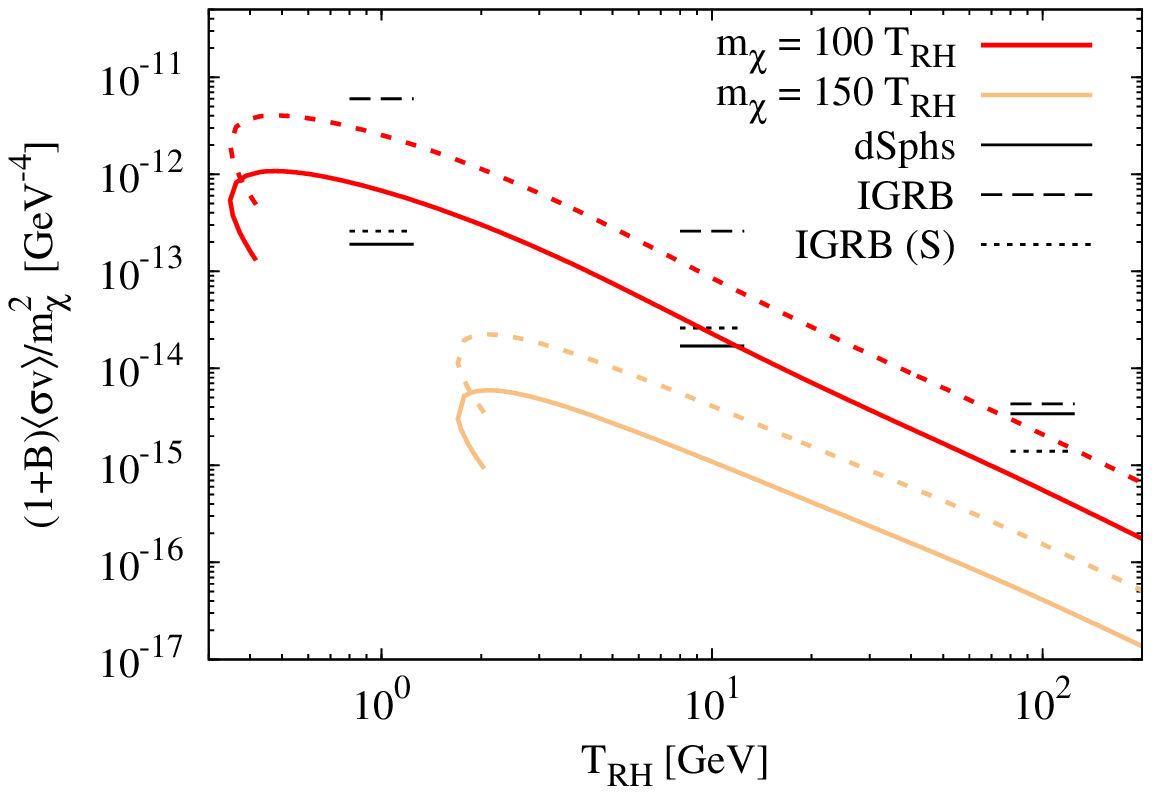}
 }
\caption{The ratio $\sigv/\mdm^2$ times the relative boost factor from EMDE-generated microhalos, along with current upper bounds on $\sigv/\mdm^2$ from Fermi-LAT observations.   The curves show values of $(1+B)\sigv/\mdm^2$ and $\Trh$ that generate the observed dark matter abundance for two values of $\mdm/\Trh$ and two values of the boost factor $B$.   The solid curves have $1+B$ = 20,000 (appropriate for dSphs), and the dashed curves have $1+B$ = 75,000 (appropriate for the IGRB).   The line segments show existing upper bounds on $\sigv/\mdm^2$ for $\mdm = 100$ GeV, 1 TeV, and 10 TeV from Fermi-LAT observations of dSphs \cite{FermidSphs15} and the IGRB \cite{FermiIGRB15}.  The dotted segments labeled IGRB (S) show the upper bounds on $\sigv/\mdm^2$ that could be obtained if the astrophysical contributions to the IGRB can be removed \cite{FermiIGRB15}.  These line segments are positioned so that $\mdm = 100 \Trh$; they may be directly compared to the $\mdm = 100 \Trh$ curves.}
\label{Fig:Detect}
\end{figure}

To use these limits on the dark matter annihilation rate to constrain EMDE cosmologies, we must determine the appropriate boost factor.  Since no boost factor was included in the analysis of the dSphs, the ratio $\sigv/\mdm^2$ should be multiplied by the total boost factor $1+B$ prior to comparison with the derived upper limits on $\sigv/\mdm^2$.   The dSphs with the largest $J$ factors have masses around $10^6 M_\odot$.   As seen in the left panel of Figure \ref{Fig:Boost}, \mbox{$1+B \lsim$ 200,000} for a halo with a mass of $10^6 M_\odot$.
However, as discussed earlier, the Fermi-LAT's limited annihilation region effectively reduces the boost factor by roughly a factor of 10.  Therefore, the effective boost factor for dSphs is about 20,000 for $z_f =400$ and $f_\mathrm{tot}(z_f) = 0.6$, which corresponds to $\kcut/\krh = 40$, as seen in Figure \ref{Fig:BFtot}.  

The solid curves in Figure \ref{Fig:Detect} show 20,000$\sigv/\mdm^2$ for EMDE scenarios that generate the observed dark matter relic abundance along with the upper bounds on $\sigv/\mdm^2$ for $\mdm = 100$ GeV, 1 TeV, and 10 TeV derived from Fermi-LAT observations of dSphs.  Figure \ref{Fig:Detect} indicates that current limits on gamma-ray emission from dSphs are already capable of ruling out EMDE scenarios with $\mdm \lsim 100 \Trh$, $\Trh \lsim 10$ GeV, and $\kcut/\krh \gsim 40$, which corresponds to $\Tkds \gsim 3 \Trh$.  Although a more detailed computation of the boost factor for each individual dSph would be required to firmly establish these constraints, this preliminary estimate encourages such an undertaking.  

Figure \ref{Fig:Detect} also illustrates the potential power of the the IGRB to constrain EMDE scenarios.  Since the Fermi-LAT analysis of the IGRB included a boost factor from standard subhalos with masses greater than $\sim$$10^{-6} M_\odot$, the ratio $\sigv/\mdm^2$ should be multiplied by the relative boost factor 
\beq
\frac{1+B}{1+B_s(M_\mathrm{min} = 10^{-6} M_\odot)} \simeq \frac52 \frac{1+B}{1+B_s(M_\mathrm{min} = 10^{-12} M_\odot)}.
\eeq
If $z_f = 400$ and $f_\mathrm{tot}(z_f) = 0.6$, then the right panel Figure \ref{Fig:Boost} tells us that the appropriate boost factor is $5/2 \times$30,000=75,000.  In Figure \ref{Fig:Detect}, this boost factor corresponds to the dashed curves.   Although 75,000$\sigv/\mdm^2$ falls below the current IGRB bounds on the annihilation rate for all interesting values of $\mdm/\Trh$, the IGRB may be able to constrain scenarios with $\kcut/\krh \gsim 40$ because microhalos form at redshifts higher than 400 in these scenarios.  The boost factor may be significantly greater than 75,000 if these early-forming microhalos survive as subhalos within the microhalos present at a redshift of 400.  Even if the microhalos at $z_f = 400$ are devoid of substructure, accounting for the earlier formation times of the smaller microhalos present at that redshift may increase the boost factor by the factor of two required to reach the observational constraints.  Figure \ref{Fig:Detect} also shows that the Fermi-LAT observations of the IGRB are easily sensitive enough to detect gamma rays from dark matter annihilations in EMDE scenarios with $\mdm \simeq 100\Trh$ if the astrophysical background can be removed.

Since the ratio $\sigv/\mdm^2$ for scenarios that generate the observed relic abundance decreases rapidly as $\mdm/\Trh$ increases (see Figure \ref{Fig:AnnFactor}), it is unlikely that gamma-ray observations will be able to constrain EMDE scenarios with $\mdm/\Trh \gsim 200$.  As shown in Figure \ref{Fig:Detect}, even $\mdm/\Trh = 150$ lies well within current observational bounds on the dark matter annihilation rate.  Furthermore, the large boost factors ($1+B \gsim$ 20,000) required to bring the annihilation rate above current observational bounds cannot be obtained if $z_f \lsim 100$.  Since microhalos do not form before $z \simeq 100$ if $\kcut/\krh = 10$, as shown in Figure \ref{Fig:BFtot}, only scenarios with $\kcut/\krh \gsim 20$ can generate large enough boost factors to saturate current observational constraints.  

Although the estimate of the boost factor developed in this section indicates that $\kcut/\krh \gsim 40$ is required to obtain boost factors greater than $10^4$, it is important to remember that the boost factor may greatly exceed this estimate.  It is possible that the boost from later-forming microhalos, which is neglected when we take $z_f = 400$, is sufficient to make $1+B \simeq 10^4$ with $\kcut/\krh \simeq 20$.  In light of this uncertainty, current observations may be able to constrain EMDE scenarios with $\kcut/\krh \gsim 20$, which corresponds to $\Tkds \gsim 2.1 \Trh$ for $\mdm \simeq 100 \Trh$.  

Any improvement in the observational constraints on the dark matter annihilation rate could extend the reach of these observations to EMDE scenarios with larger $\mdm/\Trh$ ratios or lower decoupling temperatures.  Four more years of Fermi-LAT observations of the dSph sample used in Ref. \cite{FermidSphs15} is expected to moderately strengthen the upper bounds on $\sigv/\mdm^2$ \cite{GKW14}.  Including the ultra-faint dSph candidates that were recently discovered in the Dark Energy Survey \cite{DES15a,DES15b} could provide a more dramatic improvement when the dark matter content of these systems is established \cite{FermiDES15}.  Finally, a better understanding of the astrophysical contributions to the IGRB would enable their removal, which would make the IGRB limits on dark matter annihilation far more sensitive, as seen in Figure \ref{Fig:Detect} \cite{FermiIGRB15}.

\section{Summary and Discussion}
\label{sec:con}

A transient period of effective matter domination prior to BBN is a generic feature of inflation theories and string theories that include gravitationally coupled moduli fields.  Moreover, the lightest modulus is expected to dominate the energy density of the Universe at temperatures higher than a few hundred GeV if supersymmetry provides even a partial solution to the electroweak hierarchy problem \cite[e.g.][]{KSW15}.  This delayed onset of radiation domination dramatically alters the relationship between the properties of the dark matter particle and its present-day density \cite{KT90, CKR99, GKR01, FRS03, Pallis04, GG06, DIK06, GGSY06, Drewes14, RTT14, KKN15}.  Consequently, limits on the dark matter annihilation cross section cannot rule out a thermal origin for dark matter as long as the reheat temperature is unknown.  Unfortunately, the period between inflation and BBN is difficult to probe; nonthermal expansion histories are fully consistent with the primordial abundance of light elements and cosmological observations of large-scale structure provided that the reheat temperature is at least 3 MeV \cite{KKS99, KKS00, Han04, IKT05, IKT07,dBPM08}.   

Microhalos offer a new window into the early Universe because they form from density perturbations that enter the cosmological horizon prior to BBN.  Earlier analyses of perturbation evolution during reheating \cite{ES11, BR14, FOW14} have established that dark matter density perturbations grow linearly with the scale factor prior to reheating if the dark matter is a decay product of the scalar field that dominates the Universe during the EMDE.  In this work, I have extended this investigation to include dark matter that is thermally produced during the EMDE.  Although thermal and nonthermal dark matter perturbations evolve differently at early times, the two scenarios converge after dark matter thermally decouples.  Furthermore, the final amplitude of the perturbations does not depend on whether or not the dark matter reaches thermal equilibrium prior to decoupling.  Consequently, the post-reheating matter power spectrum for both the freeze-in and freeze-out production mechanisms matches the power spectrum derived for nonannihilating nonthermal dark matter in Ref. \cite{ES11}.  In all cases, the linear growth of matter perturbations during the EMDE significantly enhances the amplitude of small-scale dark matter density perturbations.

The velocity dispersion of the dark matter particles and their elastic scatterings with relativistic particles can suppress small-scale perturbations and erase the perturbations that grow during an EMDE.  This is a crucial concern for nonthermal dark matter because free-streaming erases the perturbations that grow during an EMDE if the dark matter particles are relativistic at reheating \cite{ES11, FOW14}.  If dark matter is produced thermally, however, it may kinetically decouple during the EMDE.  In this case, the cut-off in the matter power spectrum from free-streaming and elastic scatterings does not suppress all the scales that are affected by an EMDE.  Furthermore, dark matter particles decouple earlier and cool faster during an EMDE than they would in a radiation-dominated universe, so a small relative difference between the standard decoupling temperature and the reheat temperature leads to a much larger relative difference between the cut-off scale and the horizon size at reheating.  If dark matter kinetically decouples at a temperature $\Tkds$ in a radiation-dominated universe, then $\Tkds \gsim 2\Trh$ is sufficient to preserve the enhanced inhomogeneity generated during an EMDE.  Consequently, an EMDE significantly enhances the microhalo population predicted by the Press-Schechter mass function if dark matter thermally and kinetically decouples prior to reheating.  For $\Tkds \gsim 2\Trh$, most of the dark matter is bound into microhalos at a redshift of 100.  These early-forming microhalos are far denser than the microhalos that form in purely thermal cosmologies, so they should survive outside the innermost region of a larger halo.   

A thermal origin for dark matter provides a means to detect these microhalos: dark matter annihilations within their dense cores will boost the overall annihilation rate within their host halos.  Unfortunately, this substructure does not automatically lead to an overall enhancement of the annihilation rate because a smaller value of the annihilation cross section is required to generate the observed dark matter abundance during an EMDE.  If the dark matter particle freezes out during an EMDE and $\mdm \gsim 100 \Trh$, the ratio $\sigv/\mdm^2$ required to generate the observed dark matter density is at least three orders of magnitude smaller than it is for a 100 GeV particle that freezes out during radiation domination.   Freeze-in scenarios demand even smaller annihilation cross sections for the same particle masses and reheat temperatures.  The resulting reduction in the dark matter annihilation rate is so severe that annihilations of thermal relics generated during an EMDE were thought to be undetectable, but an EMDE's effect on the microhalo population could bring these scenarios within reach of current gamma-ray observations.

To estimate the substructure boost factor from an EMDE, Press-Schechter mass functions were used to predict the fraction of dark matter that is contained in microhalos at a certain redshift.  All the microhalos at that redshift were assumed to have NFW profiles with $c=2$.  The density profile within the scale radius was assumed to be unaltered by the microhalos' subsequent absorption into their host halos, except in the innermost region of the host halo where the microhalos' tidal radii are less than twice their scale radii.  The microhalos within this region are most likely destroyed, so they do not contribute to the boost factor.  Outside this region, the distribution of the microhalos was assumed to follow the dark matter density profile of the host halo, which was taken to be an NFW profile with the concentration-mass relation proposed in Ref. \cite{SP14}.  

 The resulting boost factor is largely insensitive to the reheat temperature, but it depends very strongly on the redshift at which the microhalo mass function is evaluated.  If 90\% of the dark matter is contained in microhalos at a redshift of 50 (as is predicted for $\Tkds \gsim 3\Trh$), the relative boost factor from an EMDE is 100, which is not large enough to overcome the suppression of the annihilation cross section.  The evolution of the Press-Schechter mass function following an EMDE is strongly hierarchical, however, and microhalos are also common at very high redshift ($z \gsim 400$ for $\Tkds \gsim 3\Trh$).  If these early-forming microhalos survive as subhalos within larger microhalos, then the relative boost factor from an EMDE can exceed 30,000.  This boost factor is sufficient to make the dark matter annihilation rate within dSphs exceed the limits established by Fermi-LAT \cite{FermidSphs15} if $\mdm \simeq 100 \Trh$ and $\Trh \lsim 10$ GeV.  It would be interesting to search for dark matter candidates that have the masses and cross sections required to realize this scenario, as these models may already be in tension with observations.  The IGRB \cite{FermiIGRB15} may also constrain these scenarios; increasing the estimated boost factor by a factor of two would be sufficient to saturate current lGRB bounds on the dark matter annihilation rate.  Improving the characterization of the microhalo population could easily increase the boost factor by this amount.

The primary source of uncertainty in the EMDE boost factor is the internal structure and substructure content of the microhalos that form from the enhanced small-scale perturbations.  The Press-Schechter formalism only provides a prediction for the mass function; it cannot determine if the numerous microhalos present at a redshift of 400 survive within the larger microhalos that contain most of the dark matter at a redshift of 50.  Numerical simulations of microhalo formation in EMDE cosmologies, similar to the simulations of microhalo formation that have already been done for thermal histories \cite{DMS05, IME10, AD12, Ishiyama14}, are required to determine the internal structure of the microhalos.  The extremely hierarchical nature of microhalo formation following an EMDE facilitates these simulations because EMDE-generated microhalos form much earlier than significantly larger halos.  Therefore, it should be possible to extract a microhalo mass function from simulations in EMDE cosmologies, providing a way to check the predictions of the Press-Schechter formalism.  

The annihilation boost factor from an EMDE is also strongly dependent on the dark matter decoupling scale; a cut-off in the small-scale power spectrum determines when the first microhalos form.  In EMDE cosmologies without a small-scale cut-off, all of the dark matter is bound into microhalos at arbitrarily high redshift, which could lead to very high annihilation rates.  In this work, the small-scale power spectrum was assumed to be exponentially suppressed on scales smaller than the free-streaming length and the horizon size at kinetic decoupling.  This treatment is based on analyses of dark matter decoupling during radiation domination \cite[e.g.][]{Bertschinger06}.  However, the radiation perturbation evolves very differently during an EMDE \cite{ES11}, so it is possible that elastic scatterings between dark matter particles and relativistic particles do not have the same effect on perturbations as they do during radiation domination.  An analysis of perturbation evolution through kinetic decoupling during an EMDE \cite{IEinprep} will determine the exact relationship between the cut-off scale and the elastic scattering cross section and may extend the reach of gamma-ray observations beyond the $\Tkds \gsim 2\Trh$ limit established here.  

If gamma rays from dark matter annihilations are detected, we must disentangle the dark matter particle's properties from the potential boost factor from an enhanced microhalo population.  Fortunately, the impact of an EMDE on the spatial variation of the dark matter annihilation rate cannot be mimicked by simply increasing the annihilation cross section.  Since microhalos are not expected to survive within the innermost kiloparsec of the Galaxy, emission from the Galactic center will not be affected by an EMDE.  Therefore, a distinctive signature of an EMDE would be an annihilation signal observed in both dSphs and the IGRB with no corresponding signal from the Galactic center.  Furthermore, if the boost factor from an EMDE is significant, then the gamma-ray signal from dark matter annihilations within halos will be proportional to the number density of microhalos, which will scale with the density of the halo and not its square.  As a result, an EMDE can make the gamma-ray emission from dark matter annihilations resemble the emission from decaying dark matter outside of the halo's innermost region, but the strength of the signal from the halo center would be inconsistent with this interpretation.   Finally, an enhanced microhalo population would also affect the gamma-ray angular power spectrum \cite{2013MNRAS.429.1529F}, which may provide an additional way to determine if the annihilation rate has been boosted by an EMDE.  

An EMDE widens the field of dark matter candidates; particles with small annihilation cross sections that would be overproduced in a radiation-dominated universe are viable dark matter candidates if they freeze out before reheating.  I have shown that the effect of an EMDE on small-scale perturbations in these scenarios provides an additional observational signature that can be used to constrain these models.  If dark matter kinetically decouples before reheating, the growth of perturbations during the EMDE leads to an abundance of microhalos that boosts the dark matter annihilation rate and alters its spatial variation.  Therefore, both the absence of gamma rays from dark matter annihilation and the properties of a signal detected in the future can provide a new window into the early Universe, increasing our understanding of inflation, reheating, and the origins of dark matter.  

\acknowledgments 
I thank Scott Watson and Kris Sigurdson for several useful discussions over the course of this investigation.  I also thank Nick Priore for providing a valuable review of dark matter freeze-out abundances during an EMDE.   Finally, I thank Daniel Grin for his comments on this manuscript.  This work was supported by NSF Grant No. PHY-1417446.

\appendix
\section{Derivation of the Perturbation Equations}
\label{sec:perts}
The equations that govern the evolution of the perturbations are derived by perturbing the covariant form of the energy-transfer equations given in Eq.~(\ref{bkgd}).  Ref. \cite{LM07} used this method to derive equations for the evolution of density perturbations in annihilating dark matter on superhorizon scales during reheating in the curvaton scenario.  The same approach was used in Ref. \cite{ES11} to obtain the evolution equation for density and velocity perturbations on all scales for nonthermal dark matter.  Refs. \cite{BR14, FOW14} recently extended the analysis of Ref. \cite{ES11} to include dark matter annihilations.  In this appendix, I review this derivation and apply it to the case that dark matter is not generated during scalar decays.

The oscillating scalar field, the radiation, and the dark matter are all treated as perfect fluids with energy-momentum tensors
\beq
T^{\mu\nu}= (\rho+p)u^\mu u^\nu + p\, g^{\mu\nu},
\eeq
where $\rho$ and $p$ are the fluid's density and pressure, respectively, and $u^\mu \equiv \drm x^\mu/\drm\lambda$ is its four-velocity.  The dark matter and the oscillating scalar fields are both treated as pressureless fluids, while the radiation has $p = \rho/3$.  Since the scalar field decays into radiation and the dark matter can self-annihilate into relativistic particles, these three fluids exchange energy, as described in Eq.~(\ref{bkgd}).  This energy exchange can be expressed covariantly as
\beq
\nabla_\mu \left( ^{(i)}{T^{\mu}}_{\nu}\right) = Q^{(i)}_{\nu},
\label{cons}
\eeq
where $i$ denotes the individual fluids.   In the absence of spatial variations,
\begin{align}
\nabla_\mu\left( ^{(i)}{T^{\mu}}_{0}\right) &= - \dot{\rho_i} - 3H(\rho_i +p_i),\\
\nabla_\mu\left( ^{(i)}{T^{\mu}}_{j}\right) &= 0,
\end{align}
where a dot denotes differentiation with respect to proper time.  It is also useful to note that $T_{\mu\nu}u^\mu = - u_\nu \rho$ and $T_{\nu\lambda}T^{\lambda \beta}u_\beta = u_\nu \rho^2$.  If the fields are homogeneous, then $u_0 = -1$ and $u_i = 0$.

It follows from Eq.~(\ref{bkgd}) that
\begin{subequations} 
\label{Qexp} 
\begin{align}
Q^{(\phi)}_{\nu} &= {T^{(\phi)}_{\mu\nu}}u_\phi^\mu\gam =  - u^{(\phi)}_\nu \rho \gam,\\
Q^{(\mathrm{r})}_\nu &= - \, Q^{(\phi)}_{\nu} + L_\nu,\\
Q^{(\chi)}_\nu &= - L_\nu,
\end{align}
\end{subequations}
where
\begin{align}
L_\nu &\equiv \frac{\sigv}{\mdm} \left[^{(\chi)}T_{\nu\lambda}\,^{(\chi)}T^{\lambda \beta} - ^{(\chi,\mathrm{eq})}T_{\nu\lambda}\, ^{(\chi,\mathrm{eq})}T^{\lambda \beta}\right]u^{(\chi)}_\beta \nonumber \\
&= \frac{\sigv}{\mdm}(\rhom^2 - \rhoeq^2)u^{(\chi)}_\nu
\end{align}
captures the energy exchanged between the radiation bath and the dark matter through annihilation and pair production. 
In this three-fluid model of reheating, 
\beq
Q^{(\phi)}_{\nu}+Q^{\mathrm{(r)}}_{\nu}+Q^{\mathrm{(\chi)}}_{\nu} = 0,
\eeq
as required by the conservation of energy and momentum.

The perturbation equations are obtained by evaluating Eq.~(\ref{cons}) with the perturbed FRW metric 
\beq
\drm s^2 = -(1+2\Psi)\drm t^2 + a^2(t)\delta_{ij}(1+2\Phi)\drm x^i \drm x^j
\label{metric}
\eeq
and perturbations in the density of each fluid: \mbox{$\rho_{i}(t, \vec{x}) = \rho_i^0(t)[1+\delta_i(t, \vec{x})]$}.  I also introduce perturbations to the four-velocity of each fluid: $u_0 = -(1+\Psi)$ and $u_{j{(i)}}= a^2 \delta_{kj}v_{(i)}^k$, where $v_{(i)}^j \equiv \drm x^j/\drm t$ is the peculiar fluid velocity of the $i$th fluid in comoving coordinates.  It follows that
\begin{align}
Q^{(\phi)}_{0} &= \gam\rhos^0(1+\dels+\Psi),\\
Q^{(\phi)}_{j} &= -\gam\rhos^0 a^2 \delta_{kj} v_\phi^k,
\end{align}
to first order in the perturbations.  Note that $Q^{(\phi)}_{j}$ is first order in the perturbations, while $Q^{(\phi)}_{0}$ has both a zeroth-order component $[Q^{(\phi),(0)}_{0}=\gam\rhos^0]$  and a first-order component $[Q^{(\phi),(1)}_{0}=\gam\rhos^0(\dels+\Psi)]$.  In addition,
\begin{align}
L_0 &= -\frac{\sigv}{\mdm} \big[ (\rhom^0)^2(1+2 \delm +\Psi) \nonumber \\
&\quad\quad\quad\quad\,\,\,-(\rhoeq^0)^2(1+2 \deleq +\Psi)\big],\\
L_j &= \frac{\sigv}{\mdm}a^2 \delta_{kj} v_\chi^k \left[ (\rhom^0)^2- (\rhoeq^0)^2\right],
\end{align}
where $\deleq$  is the perturbation in the equilibrium density of the dark matter defined in Eq.~(\ref{deleq}).  Like $Q^{(\phi)}_{\mu},$ $L_j$ is a first-order quantity, while $L_0$ has both a zeroth-order component $L^{(0)}_0 = -(\sigv/\mdm) \left[ (\rhom^0)^2- (\rhoeq^0)^2\right]$, and a first-order component 
\beq
L^{(1)}_0 = -\frac{\sigv}{\mdm}  \left[ (\rhom^0)^2(2\delm+\Psi)- (\rhoeq^0)^2(2\deleq+\Psi)\right].
\eeq  

The $\nu=0$ component of Eq.~(\ref{cons}) implies that each fluid obeys the equation
\beq
\frac{\drm\delta}{\drm t}+ (1+w)\frac{\theta}{a} + 3 (1+w)\frac{\drm\Phi}{\drm t}= \frac{1}{\rho^0}\left[Q_0^{(0)} \delta -Q_0^{(1)}\right],
\label{energyeqn}
\eeq
where $w \equiv p/\rho$ is the fluid's equation of state parameter, $\theta \equiv a\, \partial_i v^i$ is the divergence of the fluid's physical velocity, and $Q_0^{(0)}$ and $Q_0^{(1)}$ are the zeroth-order and first-order components of $Q_0$ for this fluid.  The divergence of the spatial components of Eq.~(\ref{cons}) implies that
\begin{align}
\frac{\drm\theta}{\drm t} + (1-3w)H\theta + \frac{\nabla^2 \Psi}{a} + &\frac{w}{1+w}\frac{\nabla^2 \delta}{a} \nonumber \\
&= \frac{1}{\rho^0}\left[\frac{\partial_i Q_i}{a(1+w)} + Q_0^{(0)}\theta\right],
\label{momentumeqn}
\end{align}
where
\begin{align}
\frac{\partial_i Q^{(\phi)}_i}{a(1+w)} &= - \gam\rhos^{(0)} \frac{\thes}{(1+w)}\\
\frac{\partial_i L_i}{a(1+w)} &= \frac{\sigv}{\mdm}\frac{\them}{1+w}\left[ (\rhom^0)^2- (\rhoeq^0)^2\right]  \nonumber\\
&= - \frac{\them}{1+w} L^{(0)}_0.
\end{align}

Applying Eqs.~(\ref{energyeqn}) and (\ref{momentumeqn}) to the scalar, radiation, and dark matter perturbations yields
\allowdisplaybreaks
\begin{widetext}
\begin{subequations}
\begin{align}
\frac{\drm \dels}{\drm t}+\frac{\thes}{a} + 3\frac{\drm \Phi}{\drm t} &= -\gam \Psi, \label{dels}\\
\frac{\drm \thes}{\drm t} +H\thes +\frac{\nabla^2 \Psi}{a}&=0,\\
\frac{\drm \delm}{\drm t}+\frac{\them}{a} + 3\frac{\drm \Phi}{\drm t} &= \frac{\sigv}{\mdm\rhom^0}\left[-\Psi\{(\rhom^0)^2- (\rhoeq^0)^2\} - (\rhom^0)^2\delm + (\rhoeq^0)^2(2\deleq-\delm)\right], \label{delm} \\
\frac{\drm \them}{\drm t} +H\them +\frac{\nabla^2 \Psi}{a}&= 0, \label{them} \\
\frac{\drm \delr}{\drm t}+\frac{4}{3}\frac{\ther}{a} + 4\frac{\drm \Phi}{\drm t} &= \gam \frac{\rhos^0}{\rhor^0} (\dels-\delr+\Psi) + \frac{\sigv}{\mdm\rhor^0}\left[\Psi\{(\rhom^0)^2- (\rhoeq^0)^2\} +(\rhom^0)^2(2\delm-\delr) - (\rhoeq^0)^2(2\deleq-\delr)\right],\\
\frac{\drm \ther}{\drm t} + \frac{\nabla^2 \Psi}{a} + \frac{1}{4}\frac{\nabla^2 \delr}{a} &=\gam \frac{\rhos^0}{\rhor^0} \left(\frac{3}{4}\thes -\ther\right) - \frac{\sigv}{\mdm\rhor^0} \left[ (\rhom^0)^2- (\rhoeq^0)^2\right]\left(\ther - \frac{3}{4} \them\right).
\end{align}
\label{perts}%
\end{subequations}
\end{widetext}
The Einstein equations for the gravitational potentials $\Psi$ and $\Phi$ imply
\beq
\frac{\nabla^2 \Phi}{a^2} +3H\left(H\Psi-\frac{\drm\Phi}{\drm t}\right)=-4\pi G \left(\rhos^0\dels+\rhor^0\delr+\rhom^0\delm\right). 
\eeq
Since the radiation fluid is tightly coupled, the anisotropic stress is negligible, and the Einstein equations demand that \mbox{$\Phi = - \Psi$}.  

\section{Perturbation Initial Conditions}
\label{sec:ics}

The solution to Eq.~(\ref{perts}) is obtained by expressing the variables as functions of $a$ instead of $t$ and then numerically integrating the resulting set of coupled differential equations for a single plane-wave perturbation mode with wave number $k$ from $a = 1$ to some value of $a > \adec = (\gam/H_1)^{-2/3}$, where $H_1 = H(a=1)$ is the initial value of the Hubble parameter \cite{ES11, BR14, FOW14}. The integration begins when the mode is outside the cosmological horizon: $k \ll aH_1$.  The addition of dark matter annihilations does not affect the evolution of the perturbations in the scalar field, so the evolution of the scalar perturbations and the gravitational potential $\Phi$ during the EMDE is the same that as derived in Ref. \cite{ES11}:
\begin{subequations} 
\begin{align}
\Phi &= \Phi_0, \\
\dels&= 2 \Phi_0 + \frac23 \tilk^2 \Phi_0 a,  \label{scalarics:dels}\\
\frac{\thes}{H_1}  &= - \frac23 \tilk^2 \Phi_0 \sqrt{a},
\end{align}
\label{scalarics}%
\end{subequations}
where $\tilk\equiv k/H_1$.  Since the number of relativistic particles created in scalar decays greatly exceeds the number of relativistic particles created or destroyed via dark matter annihilations, adding dark matter annihilations also does not affect the evolution of the perturbations in the radiation.  For superhorizon modes \cite{ES11},
\begin{subequations}
\begin{align}
\delr &= \Phi_0 + \frac{46}{63} \tilk^2 \Phi_0 a,\\
\frac{\ther}{H_1} &= - \frac23 \tilk^2 \Phi_0 \sqrt{a}.
\end{align}
\label{radics}
\end{subequations}

The background equations given by Eq.~(\ref{bkgd}) apply to both relativistic and nonrelativistic dark matter, but the perturbation equations given by Eq.~(\ref{perts}) are only applicable to nonrelativistic dark matter.  Therefore, the numerical integration of Eq.~(\ref{perts}) must start at a temperature such that $T/\mdm <1$.  For the freeze-out scenario, however, we can only capture the evolution of the perturbations during freeze-out if the initial temperature exceeds the freeze-out temperature.  Since $T_f \lsim \mdm/5$, I start the integration of Eq.~(\ref{perts}) when $\mdm/T = 3$ for values of $\sigv$ that allow the dark matter to reach thermal equilibrium.  While the dark matter is in thermal equilibrium, the individual terms on the rhs of Eq.~(\ref{delm}) are much larger than the terms on the lhs, so the rhs terms must nearly sum to zero. If $\rhom \simeq \rhoeq$, the rhs of Eq.~(\ref{delm}) vanishes only if 
\begin{subequations}
\beq
\delm = \deleq,
\eeq
which establishes the initial condition for $\delm$.  Furthermore, this suite of initial conditions for $\delm$, $\delr$, and $\dels$ forms set of adiabatic perturbations because $H(\rho_i/\dot{\rho}_i)\delta_i$ is the same for all three components.  

It is less clear what the appropriate initial condition for $\delm$ is if the dark matter never reaches thermal equilibrium.  For these values of $\sigv$, the value of $\delm$ on superhorizon scales is chosen to make the perturbations adiabatic.  Equations (\ref{scalarics}) and (\ref{radics}) already ensure that $H \dels (\rhos/\dot{\rho}_\phi) = H \delr (\rhor/\dot{\rho}_r)$ during the EMDE, since $\rhos \propto a^{-3}$ and $\rhor \propto a^{-3/2}$ prior to reheating.  The dark matter perturbation preserves adiabaticity if
\beq
\delm = -\frac13 \dels \frac{a\rhom'(a)}{\rhom(a)},
\eeq
where $\rhom'(a)$ is obtained by solving Eq.~(\ref{bkgd}) and numerically differentiating $\avgE \nchi$ with respect to $a$.  Since the perturbation equations assume that the dark matter is nonrelativistic, it is best to start the numerical integration of Eq.~(\ref{perts}) in the freeze-in scenario at the lowest temperature for which the mode lies outside the horizon. 

Finally, since Eq.~(\ref{them}) is not affected by the addition of dark matter annihilations, the superhorizon evolution of $\them$ is given by 
\begin{align}
\frac{\them}{H_1} &= - \frac23 \tilk^2 \Phi_0 \sqrt{a}
\end{align}
\label{matics}%
\end{subequations}
for both freeze-in and freeze-out scenarios \cite{ES11}.   When Eqs.~(\ref{scalarics}), (\ref{radics}), and (\ref{matics}) are used to set the initial conditions for the perturbations, the modes that remain outside the cosmological horizon during the EMDE evolve to standard adiabatic initial conditions during reheating: $\delr \rightarrow 2 \Phi = 20/9 \Phi_0$, and $\delm \rightarrow 3/4 \delr$.  When these modes enter the horizon, they follow the standard evolution for density perturbations in a radiation-dominated universe: $\delr$ oscillates, while $\delm$ grows logarithmically.

\end{document}